\documentclass[
	twocolumn,
	superscriptaddress,
	preprintnumbers,
	aps,
	pre,
	amsmath,
	amssymb,
	floatfix,
	nofootinbib,
	notitlepage
]{revtex4-1}

\usepackage{graphicx}

\usepackage{color}

\begin{document}
\title{More ATP Does Not Equal More Contractility: Power And Remodelling In Reconstituted Actomyosin}

\author{Sami C.~Al-Izzi}
\email{\texttt{s.al-izzi@unsw.edu.au}}
\affiliation{School of Physics, UNSW, Sydney, NSW 2052, Australia.}
\affiliation{ARC Centre of Excellence for the Mathematical Analysis of Cellular Systems, UNSW Node, Sydney, NSW 2052, Australia.}

\author{Sedigheh Ghanbarzadeh Nodehi}
\affiliation{Centre for Mechanochemical Cell Biology and Division of Biomedical Sciences, Warwick Medical School, University of Warwick, Coventry CV4 7AL, United Kingdom.}

\author{Darius V.~K\"{o}ster}
\email{\texttt{d.koester@warwick.ac.uk}}
\affiliation{Centre for Mechanochemical Cell Biology and Division of Biomedical Sciences, Warwick Medical School, University of Warwick, Coventry CV4 7AL, United Kingdom.}

\author{Richard G.~Morris}
\email{\texttt{r.g.morris@unsw.edu.au}}
\affiliation{School of Physics, UNSW, Sydney, NSW 2052, Australia.}
\affiliation{ARC Centre of Excellence for the Mathematical Analysis of Cellular Systems, UNSW Node, Sydney, NSW 2052, Australia.}
\affiliation{EMBL Australia Node in Single Molecule Science, School of Medical Sciences, UNSW, Sydney, NSW 2052, Australia.}


\begin{abstract}
The cytoskeletal component actomyosin is a canonical example of active matter since the powerstroke cycle locally converts chemical energy in the form of adenoside triphosphate (ATP) into mechanical work for remodelling. Observing myosin II minifilaments as they remodel actin {\it in vitro}, we now report that: at high concentrations of ATP, myosin minifilaments form metastable swirling patterns that are characterised by recurrent vortex and spiral-like motifs, whereas; at low concentrations of ATP, such structures give way to aster-like patterns. To explain this, we construct the (quasi-)steady states of a polar active hydrodynamic theory of actomyosin whose ATP-scaling is obtained from a microscopic, stochastic description for the ATP-dependent binding of the heads of single myosin II minifilaments. The latter codifies the heuristic that, since the powerstroke cycle involves the unbinding of myosin II heads from actin, increases in the concentration of ATP reduce the likelihood that a given myosin II minifilament has more than one head bound simultaneously, reducing its ability to generate contractile forces and increasing the relative likelihood of processive motion. This reproduces several qualitative and some quantitative aspects of experiments, providing evidence for the central phenomenon of the theory: an ATP-dependent active contractile instability. ATP therefore controls not only the rate at which work is done--- \textit{i.e.,} the power--- but also the mode by which this occurs.
\end{abstract}

\maketitle

\section{Introduction}

Actomyosin--- the cortical meshwork of actin filaments and myosin II motors--- is a cytoskeletal component that is crucial for many cellular functions, including morphogenesis, motility and division \cite{banerjee_actin_2020,prost_active_2015,alberts_molecular_2002}. It is also a central example of active matter \cite{marchetti_hydrodynamics_2013}, since the ATPase activity of myosin II head-groups couples the hydrolysis of bound Adenosine Triphosphate (ATP) with a conformational change, locally converting chemical energy into mechanical forces via the so-called powerstroke cycle \cite{tyska_myosin_2002}.

However, whilst the rate at which work can be performed is ultimately determined by the balance between ATP and its hydrolysis products, Adenosine Diphosphate (ADP) and Phosphate (Pi), evidence from several {\it in vitro} studies increasingly suggests that the relationship between the concentration of available ATP and actomyosin behaviour is complex: ATP has been shown to influence actin meshwork pattern formation \cite{smith_molecular_2007}, rheology \cite{Mizuno2007}, actin filament breaking \cite{vogel_myosin_2013} and myosin II processivity \cite{mosby_myosin_2020}.

To better understand the interplay between remodelling and power, we combine total internal reflection fluorescence (TIRF) microscopy and interferometric scattering (iSCAT) microscopy \cite{ortega-arroyo_interferometric_2012,cole_label-free_2017} to observe actomyosin reconstituted on supported lipid bilayers (SLBs). TIRF is used to image remodelling of the actin architecture, while iSCAT--- a label-free approach that relies on the interference between reflected and scattered light from nano-objects near an interface--- is used to observe myosin II minifilaments (oligomers of up to $\sim50$ individual myosin II molecules) over long times without the deterioration and/or inactivation of myosin due to photodamage.  Notably, at intermediate-to-high concentrations of ATP, we report recurrent meta-stable actomyosin structures, such as vortex- and spiral-like motifs, where myosin II minifilaments travel along actin filaments. Such swirling and dynamic structures are in contrast with the aster-like patterns formed as the available supply of ATP depletes.

\begin{figure}[h!]
	\includegraphics[width=0.5\textwidth]{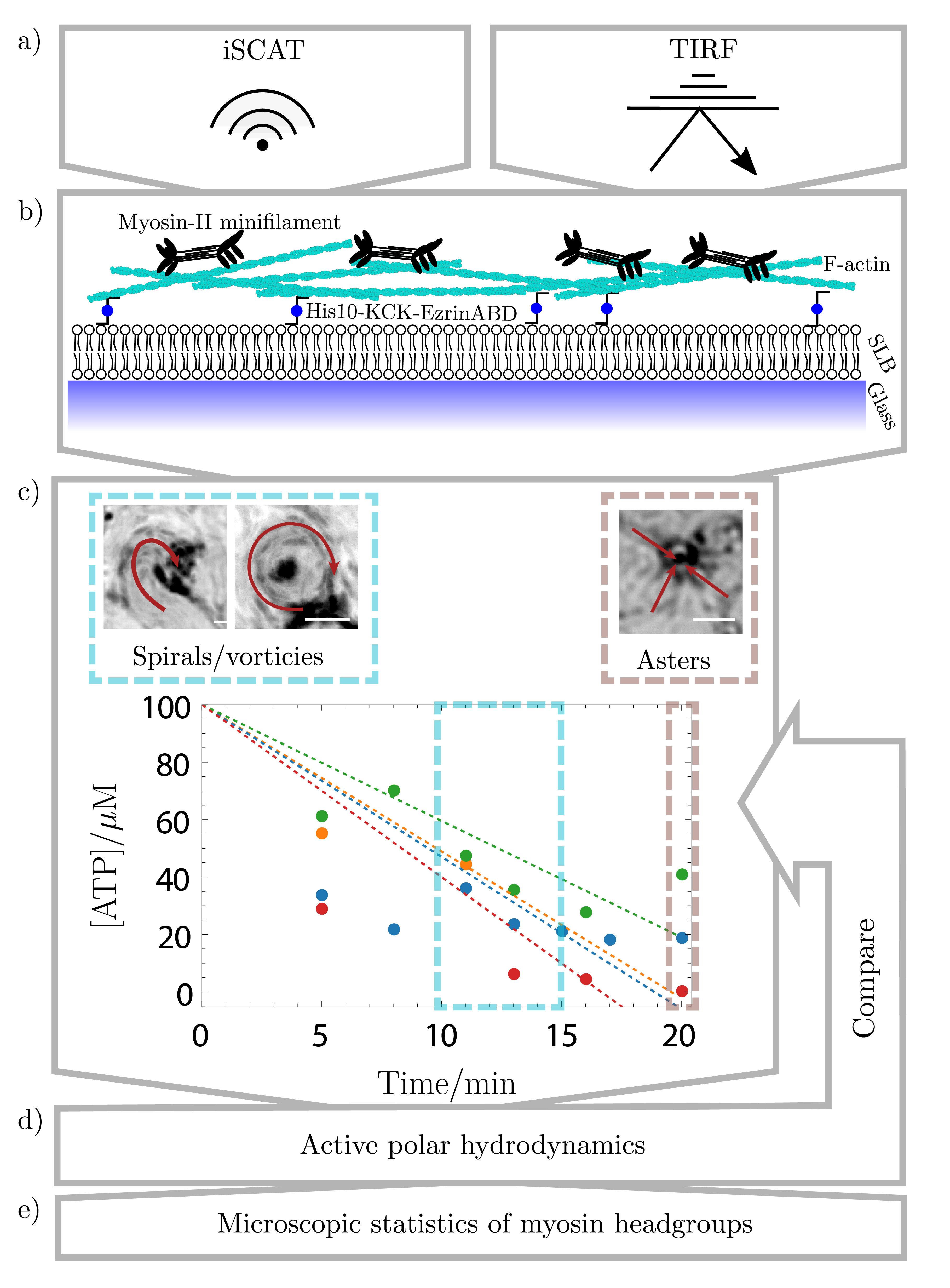}
	\caption{
		\label{fig:schematic}
		\textbf{Schematic outline of paper.} Interferometric scattering (iSCAT) and total internal reflection fluorescence (TIRF) microscopy \textbf{(a)} are used to observe actomyosin reconstituted \textit{in vitro} on a supported lipid bilayers (SLB) \textbf{(b)}. At high concentrations of ATP, myosin minifilaments form meta-stable swirling patterns that are characterised by recurrent vortex- and spiral-like motifs, where, as ATP depletes (at a rate $5.12\pm 0.40 \mu\text{M min}^{-1} $), such structures give way to aster-like patterns \textbf{(c)}. To capture this behaviour, we construct an active a polar active hydrodynamic theory \textbf{(d)} whose active coefficients take a functional form obtained from a microscopic, stochastic model of the ATP-dependent head kinetics of a single myosin II minifilament \textbf{(e)}. The resulting, composite model reproduces several qualitative and some quantitative features of our observations, and is supportive of the central aspect of our theory: an ATP-dependent contractile instability.
  }
\end{figure}

To capture the characteristic physics that underpins these dynamic and emergent patterns, we construct the (quasi-)steady states of a polar active hydrodynamic description of our \textit{in vitro} system; at both high ATP and low ATP. In order to do this, we use a microscopic, stochastic model for the bound status of the heads of a single myosin II minifilament to derive an ATP-scaling anzatz for the active coefficients in our hydrodynamic theory. This reflects the fact that, since the power stroke cycle involves the unbinding of myosin II heads from actin, increases in the concentration of ATP reduce the likelihood that a given myosin II minifilament has more than one head bound simultaneously, reducing its ability to generate contractile forces and increasing the relative likelihood of processive motion.

We compare the results with particle image velocimetry (PIV) as well as interferometric contrast and fluorescence intensity values (from iSCAT and TIRF, respectively). This shows that the composite model is consistent with several qualitative aspects of our observations but not all quantitative features. We discuss the limitations of our approach and argue that the agreement that we see reflects the central notion underpinning our model: that the nascent contractile instability--- where force-generating myosins are advected by the contractile remodelling that they induce themselves--- only occurs beneath a threshold of ATP. In other words, ATP is responsible not only for controlling the {\it amount} of work done, but also for the {\it way} it is done, resulting in a non-trivial relationship between power and remodelling in actomyosin networks.

\section{Asters, spirals and vortices revisited}

\begin{figure}
	\includegraphics[width=0.475\textwidth]{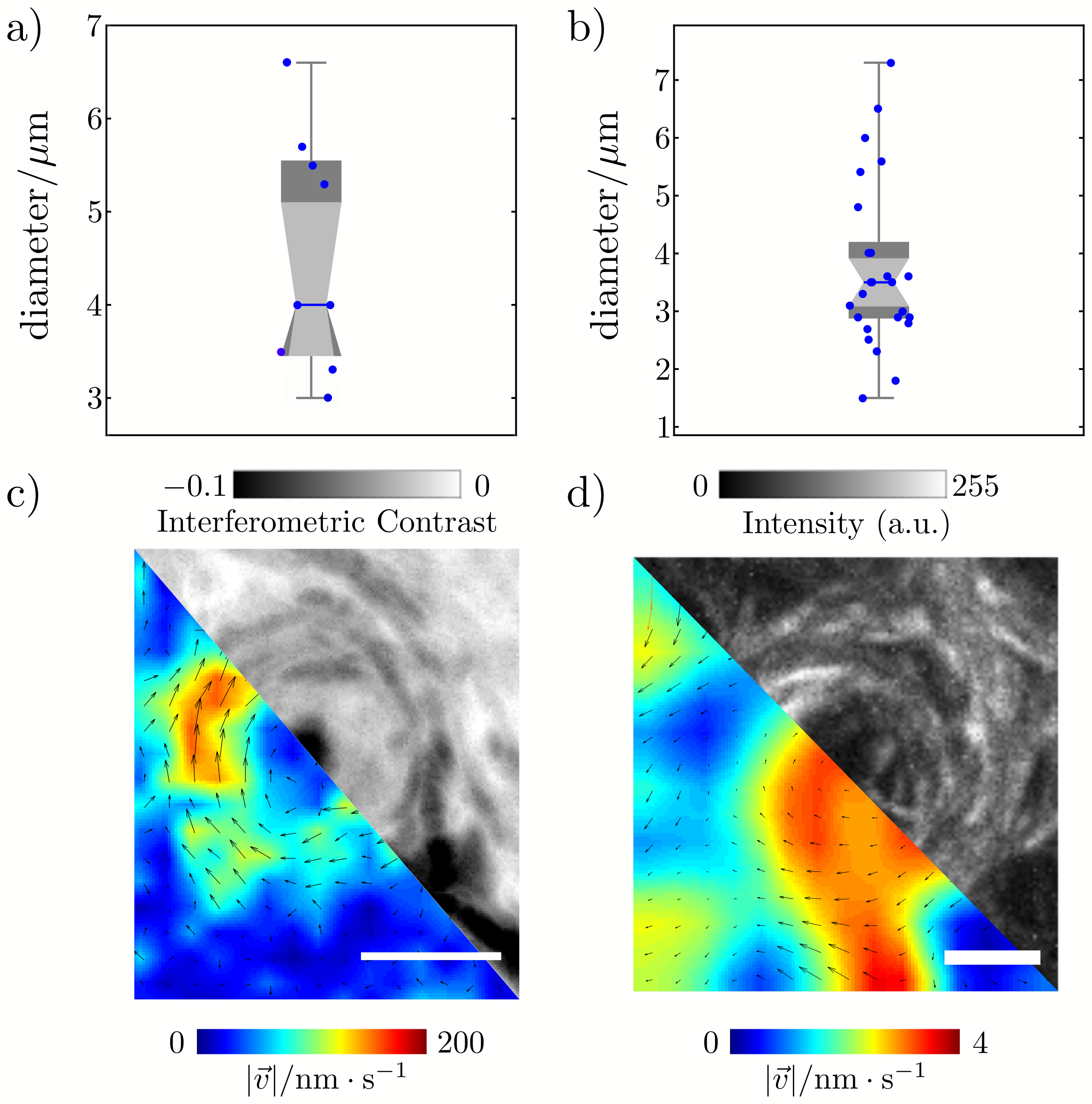}
	\caption{
		\label{fig:experimentalSetup}
		\textbf{Meta-stable vortices and spirals at intermediate-to-high concentrations of ATP.} \textbf{a)} Approximately 10 min after addition of $ 0.1\text{ mM}$ ATP, networks with intermediate length F-actin ($l_{\text{actin}} = 8 \pm 4.7 \text{ }\mu\text{m}$ \cite{meijering_design_2004}) and an approximate minifilament-to-actin filament ratio of 3:1
		displayed meta-stable spiral-like structures in iSCAT with an average diameter of $4.5\ \pm1.3$ $\mu$m. \textbf{b)} Observing a range of lengths ($l_{\text{actin}} = 9 \pm 5.5$/ $7 \pm 4.7$/ $6 \pm 4.6$ $\mu\text{m}$) of fluorescently-labelled actin via TIRF at an approximate minifilament-to-actin ratio of 1:1
		resulted in motifs with an average diameter $3.7 \pm 1.5\text{ }\mu\text{m}$.  \textbf{c)} Representative minimal intensity projections (duration $120$s, scale bar $2\mu\text{m}$) and corresponding flow fields inferred using PIV analysis for iSCAT. \textbf{d)} Representative vortex TIRF maximal intensity projection (duration $120$s, scale bar $2\mu\text{m}$) with corresponding PIV analysis. (See Appendix \ref{sec:app1} for more details). }
\end{figure}

We start with actomyosin reconstituted \textit{in vitro} on supported lipid bilayers \cite{sumino_large-scale_2012,sanchez_spontaneous_2012,koster_actomyosin_2016,koster_cortical_2016,fritzsche_cytoskeletal_2017,spudich_regulation_1971,pollard_chapter_1982}: a system previously shown to readily remodel under the action of ATP, forming asters and contractile foci under a variety of conditions \cite{koster_actomyosin_2016,koster_cortical_2016}.  Specifically, we follow \cite{mosby_myosin_2020} and use iSCAT microscopy to observe the motion of myosin II minifilaments atop a texture of intermediate length F-actin ($l_{\text{actin}} = 9 \pm 5.5$/ $7 \pm 4.7$/ $6 \pm 4.6$ $\mu\text{m}$), itself attached to fluid lipid-bilayer supported by a glass substrate (Fig. \ref{fig:experimentalSetup}{\bf a}). Our central observation is that, while aster-like contractile foci can be observed once ATP has been depleted (Vid.~1), at earlier times, when ATP concentrations are estimated to be around 10-30 $\mu$ M \cite{vogel_myosin_2013,smith_molecular_2007}, we also observe complex swirling patterns.  These are characterised by recurrent dynamical vortex and spiral-shaped motifs, where myosin II minifilaments tightly circle a void or a myosin II rich centre (Fig.~\ref{fig:experimentalSetup}d, Vid.~2-5, SM \cite{supp_al-izzi}). On rare occasions, we also see such structures merge to form a larger foci (Vid.~6). The average diameter of the circular motion was $4.5 \pm 1.3\,\mu\text{m}$, with particle image velocimetry (PIV) \cite{thielicke_flapping_2014,thielicke_pivlab_2014,garcia_fast_2011} characterising peak speeds on the order of $100\,\text{nm}\,\text{s}^{-1}$ (see SM \cite{supp_al-izzi}). The lifetimes of the motifs all exceeded $90\text{\,s}$, which is much longer than the average residence time of individual minifilaments \cite{mosby_myosin_2020}, and is indicative of a quasi-stable state whereby the incoming and outgoing fluxes of minifilaments are almost balanced.

Since the iSCAT signal of actin filaments is very low compared to the myosin signal, we separately observed fluorescently labelled actin filaments under similar conditions using total internal reflection fluorescence (TIRF) microscopy (Fig.~\ref{fig:experimentalSetup}c and Sec.~\ref{sec:app1}). Again, this revealed long-lived vortex-like motion (Fig.~\ref{fig:experimentalSetup}e, Vid.~7-8, SM \cite{supp_al-izzi}), with circular motifs of average diameter $3.7 \pm 1.5\,\mu\text{m}$ and lifetimes $>100\text{ s}$.  However, strikingly, PIV revealed that the speeds were $1-5\,\text{nm}\,\text{s}^{-1}$, significantly slower than those of the myosin II minifilaments (see SM \cite{supp_al-izzi}). In addition, actin and myosin vortices flow in the same direction, and almost all vortices are of the same handedness (29 of 34). We attribute this latter feature to the chirality of actin inducing a twist-bend coupling \cite{tanaka_super_1992,nishizaka_right-handed_1993,de_la_cruz_origin_2010}, although a detailed investigation is considered out of scope for this article.

Nominally, such observations have superficial similarities to steady-state {\it asters}, {\it spirals} and {\it vortices} reported in a generic treatment of incompressible active polar gels~\cite{kruse_asters_2004}.  However, the underlying physics of \cite{kruse_asters_2004} differs considerably from that of our system, where myosin is explicitly featured, layered over actin, and where both actin and myosin are compressible. The layering or stratification of myosin and actin has been suggested to be an important organising factor elsewhere \cite{das_stratification_2020}.  Moreover, there is a lack of evidence for noticeable differences between the elastic moduli for splay and bend, or any noticeable active spontaneous-splay stresses, which drive the rotation of spirals in \cite{kruse_asters_2004}.  As a result, the only stable steady states in our system should, in principle, be aster-like, which is in line with the actin textures that we observe on depletion of ATP (and their behaviour on ATP add-back).

To provide more quantification of the ATP consumption rate we perform a measurement of the concentration over time in a separate experiment using a fluorescent ATP reporter (Sec.~\ref{sec:app1}, Fig.~\ref{fig:schematic} c)) finding ATP consumption rates of $5.12\pm 0.40\mu\text{M min}^{-1}$, a rate in line with similar measurements on \textit{in vitro} actomyosin systems \cite{sakamoto_f-actin_2024}.

\section{Active hydrodynamics of {\it in vitro} actomyosin}

In order to explore the phenomenology that underpins our observations, we write down a nemato-hydrodynamic description of our \textit{in vitro} system (Appendix \ref{sec:app4} \& \cite{chaikin_principles_2000,marchetti_hydrodynamics_2013,kruse_asters_2004,kruse_generic_2005,husain_emergent_2017,gowrishankar_nonequilibrium_2016}).  Although this overlooks some nontrivial aspects of the rheology of transiently bound intermediate-length actin filaments (see Discussion), we argue that such a symmetry-based approach is sufficient if we restrict ourselves to long-lived steady-state and quasi-steady-state behaviours only (see comparison between theory and experiment, Section \ref{sec:comparison}).

\begin{figure*}
	\includegraphics[width=\textwidth]{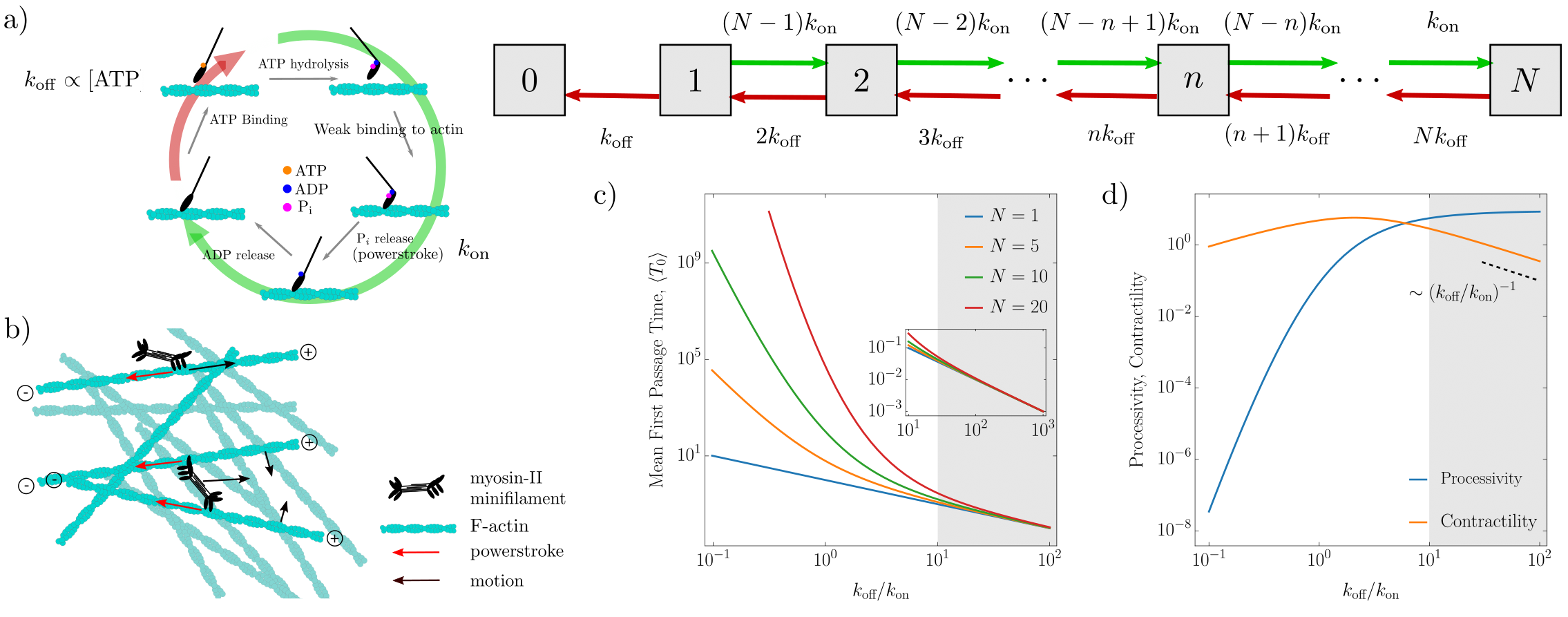}
	\caption{
		\label{fig:microscopicModel}{\bf ATP-dependent myosin II minifilament kinetics: processivity vs.~contractility}. \textbf{a)} Approximating the canonical 5-step ATP-cycle by a 2-step binding/unbinding process, the statistics of the bound status of a single myosin II minifilament's head groups can be modelled using a ladder-like state space. \textbf{b)} The power-stroke associated with the transition between singly- and doubly-bound heads leads to processive `steps' along actin filaments; additional bound heads leads to cross-linking and therefore meshwork remodelling that is contractile at the hydrodynamic scale (see \cite{koenderink_active_2009,linsmeier_disordered_2016,murrell_f-actin_2012,ideses_myosin_2013,wollrab_polarity_2019}). \textbf{c)} As $k_\text{off}\propto\text{[ATP]}$ increases, the mean first passage time, $T_0$, until the dissociation of the myosin II minifilament, converges to $\sim1/k_\text{off}$ (see inset) regardless of the total number of head groups, allowing us to identify $k\sim\langle T_0\rangle^{-1}$. \textbf{d)} The rate of processive ({\it i.e.}, $1\to 2$) and contractile ({\it i.e.}, $n\to n+1$ for $n\ne 1$) power-strokes scale as a constant and $\sim 1/k_\text{off}$, respectively, as $k_\text{off}\propto\text{[ATP]}$ increases. The shaded regions in \textbf{c)} \& \textbf{d)} correspond to the regions where we derive the analytical scaling used in our hydrodynamic model.    
		}
\end{figure*}
Since the depletion of bulk ATP is slow compared to its characteristic rate of diffusion, we adopt a quasi-static approach to ATP concentration.  Therefore, there are only two continuity equations, one each for the density of myosin II minifilaments, $\rho_m$ and the density of actin filaments, $\rho_a$. Using myosin II minifilament residence time and run length as characteristic spatio-temporal scales (Appendix \ref{sec:app4}) these take the dimensionless form
\begin{equation}
	\partial_t\rho_m + \nabla_i J^i_m =  \frac{\rho_a}{\rho_a + \rho_a^h} - k\rho_m,\label{eq:myosinDensity}
\end{equation}
and
\begin{equation}
	\partial_t\rho_a + \nabla_i J^i_a  = 0,\label{eq:actinDensity}
\end{equation}
where $J^i_m = \rho_m v^i_m$ and $J^i_a = \rho_a v^i_a$ are the components of the myosin and actin currents, respectively, and $v^i_a$ and $v^i_m$ the corresponding velocity components. The source/sink terms in the myosin equation correspond to a minifilament on-rate that depends upon actin (in a saturating form with Hill coefficient $\rho^h_a$) and a density-dependent off-rate, with $k$ the `bare' ratio of these two quantities for $\rho_a\gg \rho^h_a$.

The continuity equations (\ref{eq:myosinDensity}) and (\ref{eq:actinDensity}) are coupled by force balance conditions at the myosin-actin and actin-bilayer interfaces of our layered system (Fig.~\ref{fig:schematic}a), giving rise to constitutive relations for the currents:
\begin{equation}
J_m^i = \frac{\rho_m}{\rho_a} J_a^i - \rho_m \alpha\left(\left[ATP\right]\right)P^i - \chi_m \nabla^i\rho_m,\label{eq:myosinForceBalance}
\end{equation}
and
\begin{equation}
		J_a^i = \nabla_j\Sigma^{ij} - \chi_a \nabla^i \rho_a\text{.} \label{eq:dimensionlessActinVelcoityExpressions}
\end{equation}
Here, $\chi_a$ and $\chi_m$ are the dimensionless inverse compressibilities for actin and myosin II minifilaments, respectively, $\alpha([\text{ATP}])$ is a dimensionless coefficient that controls the ATP-dependence minifilament velocities, and $P^i$ is a polar order parameter. We assume that $\alpha([\text{ATP}])\to 1$ as $[\text{ATP}]\to \infty$ \cite{mosby_myosin_2020}. The stress tensor $\Sigma^{ij}$ is given by 
\begin{equation}
	 \Sigma^{ij} = -\zeta(\rho_m,[ATP]) g^{ij} + \frac{1}{2}\left(P^ih^j - P^j h^i\right),
			\label{eq:sigma}
\end{equation}
where $\zeta$ is an active isotropic contractility, and $g^{ij}$ is the inverse metric. The quantity $h^i=-\delta \mathcal{F}/\delta P_i$, is the so-called molecular field, which is derived from the elastic free-energy of the polarisation field \cite{de_gennes_physics_1993}. Since elastic asymmetry between bend and splay had negligible effects on our results, we assume a one-constant Frank free energy
\begin{equation}
	\mathcal{F}=\int \frac{\kappa}{2} |\nabla_i P^j|^2 \mathrm{d}A,
\end{equation}
where $\kappa$--- the Frank elastic modulus \cite{de_gennes_physics_1993}--- is assumed to be small, so as to give a short actin persistance length comparable with experiments. The molecular field is thus given by $h^i = \kappa \Delta P^i$, where $\Delta$ is the Laplacian.

The system of equations is closed by the dynamics of the polar order parameter, which is given by 
\begin{equation}
	\mathrm{D}_t P^i = \frac{h^i}{\gamma} + \nu \left[\frac{\rho_a}{\rho_a^\star}\left(1-P_jP^j\right) -\left(1+P_jP^j\right)\right]P^i,\label{eq:polarisation}
\end{equation}
where $\gamma$ controls the relaxation of $P^i$, and $\mathrm{D}_t P^i=\partial_t P^i + v^j_a\nabla_j P^i + \frac{1}{2}\left(\nabla^{i}v_{a}{}_{j}-\nabla_j v_{a}^{i}\right) P^j$ is the objective rate, or co-rotational derivative. The lyotropic term premultiplied by $\nu$ gives the Landau-de Gennes transition as a function of the order parameter density, such that for low actin density $|P^i|\to 0$, and at high densities $|P^i|\to 1$ (Appendix \ref{sec:app4}). $\rho_a^\star$ controls the density at which the nematic-isotropic transition occurs.

Activity enters explicitly into our model in three places: the ratio $k$, which governs the binding kinetics of myosin II minifilaments [Eq.~(\ref{eq:myosinDensity})]; the magnitude of isotropic contractile stresses $\zeta$ [Eq.~(\ref{eq:sigma})], and; the magnitude, $\alpha$, of the monopolar-like processive force $\propto P^i$ [Eq.~(\ref{eq:myosinForceBalance})]. For the latter, we recall that, in this dimensionless presentation, a characteristic myosin minifilament velocity has been absorbed into the definitions of various quantities, and therefore $\alpha$ is a number between 0 and 1 (Appendix \ref{sec:app4}).

In principle, each of these active quantities, as well as passive dissipative coefficients and storage moduli, can depend on variables that are not treated hydrodynamically, such as the concentration of available ATP.  Couplings of this type cannot be obtained from either (broken) symmetries or the relevant conservation principles at play, and instead must be obtained either empirically or from a microscopic model.

\section{A microscopic model of single myosin II minifilaments}

\begin{figure*}[!htp]
\includegraphics[width=\textwidth]{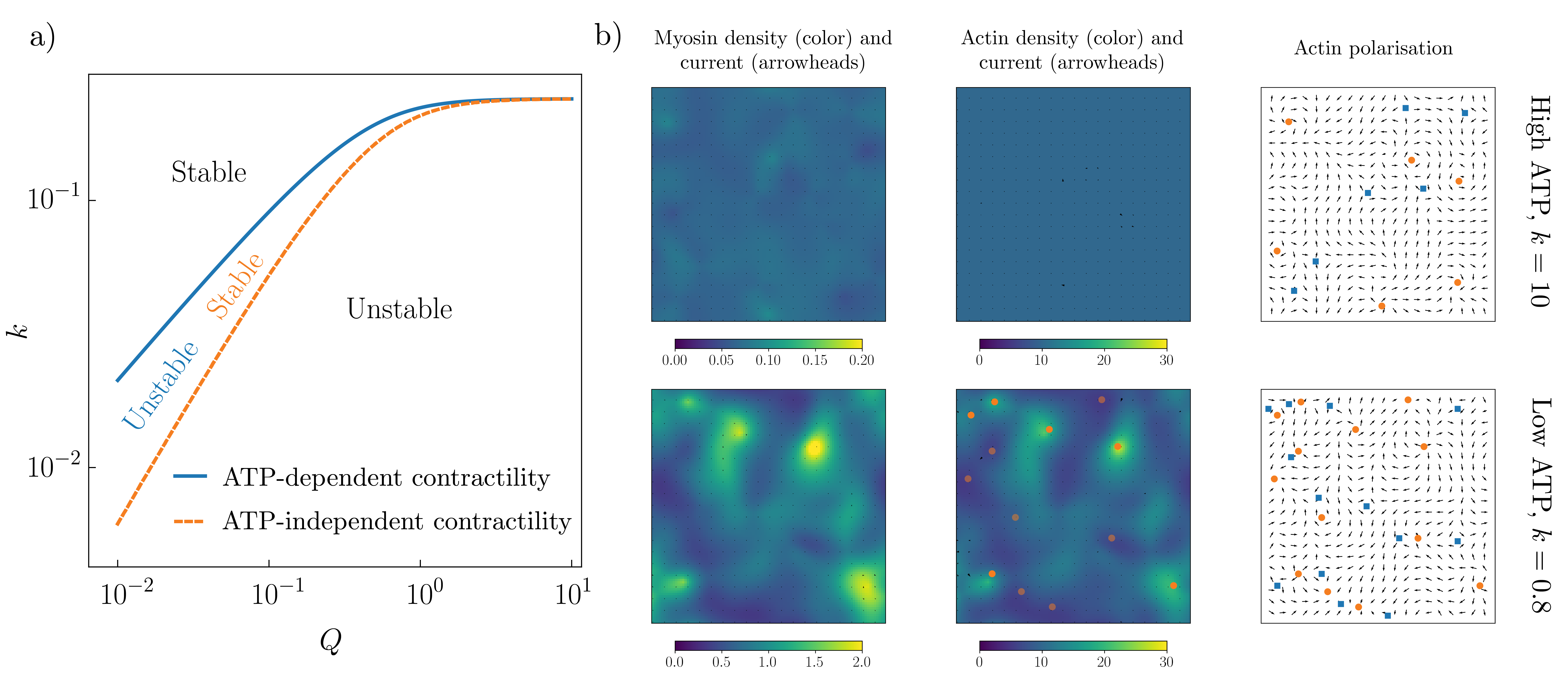}
	\caption{
		\label{fig:numericsFig} {\bf ATP-dependent criteria for an advective-contractile instability}. The microscopic model of myosin minifilament head group kinetics provides a scaling ansatz for the coefficients of our hydrodynamic theory: the myosin minifilament off-rate is directly proportional to the concentration of ATP and the active contractility is inversely proportional to the concentration of ATP. {\bf a)} The impact of this scaling can be quantified by comparing the resulting ATP-dependent stability criteria against one where active contractility is ATP independent [$Q=\sqrt{|q|^2}$ is the magnitude of the wavevector, and $k$ is the dimensionless unbinding rate of the myosin-II minifilament, {\it cf}.~Eq.~(\ref{eq:linstab})]. This shows that the ATP-dependent case has a advective-contractile instability for higher values of $k$ than would be expected in the ATP independent case. {\bf b)} Numerical solutions of the hydrodynamic equations at high ($k=10$) and low ($k=0.8$) ATP concentrations capture the effects of the stability criteria, with foci only forming in the low ATP case. We also show the corresponding actin textures, and their the topological defects ($+1$ in orange and $-1$ in blue), with the $+1$ defects overlaid on the actin density in the low ATP regime (half opacity shows defects where the plus end of actin is directed inwards, which corresponds to an outward myosin flux). The defects colocalize with the actin and myosin foci as expected. Numerics were performed from uniform densities and random (unit) polarisation initial conditions on a $100\times 100$ grid. The final time point was $t=24$ in dimensionless units, and the following parameters were used: $\kappa=0.1$, $\xi=-1$, $\chi_a=0.1$, $\chi_m=1$, $\gamma=10$, $\nu=1$, $\rho_a^\star=5$ and $\rho^h_a=5$. For further details, see Appendix \ref{sec:app4}.
		}
\end{figure*}
To motivate how the different terms in our hydrodynamic description of actomyosin scale with the concentration of available ATP, we examine a simple model that aims to capture the statistics of the bound status of the $N\approx50$ individual head groups that belong to a single myosin II minifilament.

We approximate the canonical five-step ATP cross-bridge cycle \cite{alberts_molecular_2002} using a two-step process where myosin heads bind to actin at a constant rate $k_\mathrm{on}$, and unbind with a rate $k_\mathrm{off} \sim \text{[ATP]}$ (Fig.~\ref{fig:microscopicModel}a-b). For the latter, we only assume that the off rate is monotonically increasing with ATP, and do not attempt to infer the exact proportionality with ATP. Otherwise, different heads are assumed to act independently, without considering any potential load dependence \cite{debold_slip_2005,stam_isoforms_2015} and/or mechanical cooperativity \cite{walcott_mechanical_2012,kaya_coordinated_2017}.  When $n$ heads are bound, therefore, the rate of additional binding is simply $(N-n)k_\mathrm{on}$, whilst the rate of unbinding is $n\,k_\mathrm{off}$.  This results in a ladder-like stochastic process of head binding and unbinding that inevitably ends with the dissociation of myosin II minifilaments ({\it, i.e.}, no heads bound).

In particular, the master equation associated with the above process can be solved recursively to yield the mean first passage time to dissociation of a myosin II minifilament, $\langle T_0\rangle$, having started with a single bound head (Appendix \ref{sec:app3}).  This shows that higher concentrations of ATP increase the ratio of off- to on-rates and therefore decrease residence times of myosin II minifilaments, which is qualitatively in line with more complex models of myosin II minifilaments in the low load limit \cite{erdmann_sensitivity_2016,albert_stochastic_2014} and experimental observations of the duty ratio of myosin I proteins \cite{Veigel1999a}.  Specifically, for all appreciable levels of ATP, $\langle T_0\rangle\sim k_\text{on}/k_\text{off}$, regardless of the number of heads, $N$, in a given minifilament (Fig.~\ref{fig:microscopicModel}c). Despite the fact that the minifilament population comprises oligomers of different lengths, we may therefore treat all minifilaments equally when referring to the dependence of their kinetics ({\it i.e.}, their residence times) on ATP.  This means that the quantity $k$--- the bare ratio of minifilament off-to-on-rates--- which appears in our hydrodynamic model must scale with $1/\langle T_0\rangle$, and hence $k\propto [ATP]$.   

However, extending this type of argument to active and passive coefficients requires a distribution over the {\it number} of state transitions--- {\it i.e.}, the likelihood that a given transition ({\it e.g.}, $n\to n+1$) will occur a certain number of times. This is because each ascending transition in our model involves a single power-stroke, and therefore the mean number of such transitions divided by the total minifilament residence time can be used as a proxy for the rate at which work is done, or the power. 

To understand {\it how} work is being done, we make a distinction between transitions of the type $1\to 2$, which are tantamount to taking a single processive step along the minifilament (toward the `plus' end), and all other ascending transitions, $n\to n+1$ for $n\neq 1$, which involve one or more bound heads in addition to that which performs the power-stroke. If these additional bound heads anchor myosin II to either the same actin filament or another actin filament that cannot move due to meshwork frustration, then motion stalls. Otherwise, the ATP-induced power-stroke can generate relative motion between two or more actin filaments.  On the scale of bulk hydrodynamic descriptions, the latter case manifests as predominantly contractile forces, since the response of actin filaments is asymmetric when subjected to compressive or tensile loads \cite{koenderink_active_2009,linsmeier_disordered_2016,murrell_f-actin_2012,ideses_myosin_2013}, and the dwell time of myosin II head groups at the plus end of actin filaments is abnormally long \cite{wollrab_polarity_2019}.

Using a sum-over-paths representation of our model (Appendix \ref{sec:app3}) the aforementioned proxies for the power associated with these two cases--- `processivity' and `contractility'--- can be calculated. These imply that, as ATP increases, the former converges to approximate a constant, whilst the latter scales like $1/k$ (Fig.~\ref{fig:microscopicModel}d). In the context of the hydrodynamic model, this leads us to make two assumptions. First, we assume that $\zeta$ has the form
\begin{equation}
	\zeta = \frac{\xi\rho_m}{k}.
	\label{eq:zeta}
\end{equation}
Second, we assume that the mono-polar-like force that drives the processivity of myosin is of constant magnitude, and hence $\alpha = 1$.

\section{An ATP-dependent advective-contractile instability}

\begin{figure*}[!htp]
\centering\includegraphics[width=\textwidth]{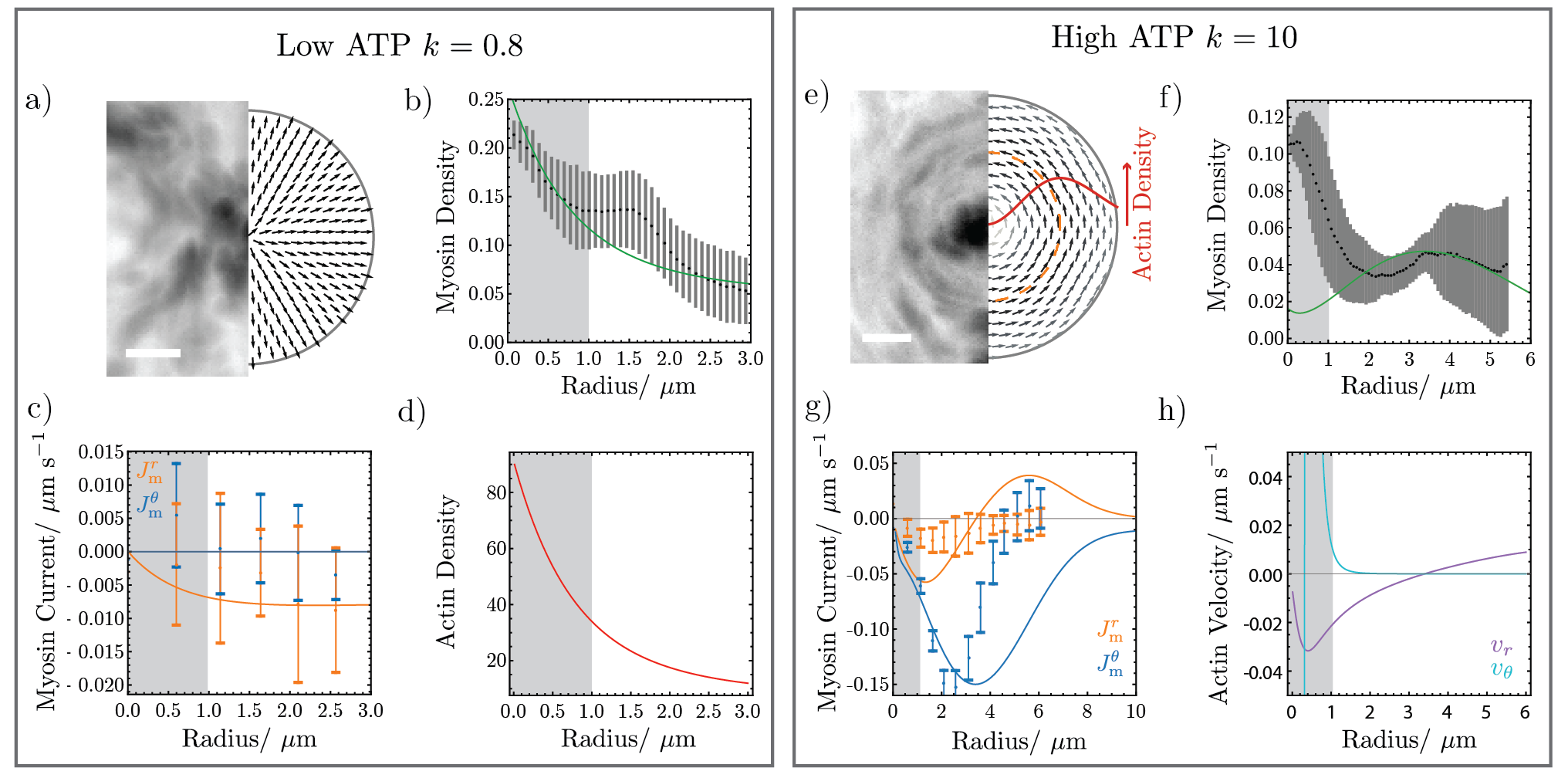}
	\caption{
		\label{fig:hydrodynamicsFig} {\bf Comparing theory and experiment}. Solving our composite theory in the cases of high and low ATP reproduces several qualitative and some quantitative features of observations outside of a small circle of radius $\sim1\mu$m, within which our theory breaks down.  \textbf{Left panel}: At low ATP, contractility is large, resulting in axis-symmetric steady-state solutions. These are asterlike actin textures on which myosins process and advect towards the centre, where the resulting build-up of myosin is balanced by an increased off-rate. 
  \textbf{a)}Left Static iSCAT microscopy image of an aster formed after depletion of ATP (20 min after addition of 0.1 mM ATP, where dark areas indicate high myosin II minifilament density).
  \textbf{a)}Right Underlying asterlike actin texture for the steady-state solution.
  \textbf{ b)} The radial myosin density profile broadly captures the angular average of the interferometric scatting intensity.
\textbf{c)} The components of the myosin current broadly agree with density weighted velocities from a PIV analysis.
\textbf{d)} The radial actin density profile (not captured by iSCAT) mirrors the functional form of the myosin density profile.
\textbf{Right panel}: At high ATP, contractility is small and the zeroth order behaviour of a perturbative solution is quasi-steady. This corresponds to a vortexlike texture with actin density peaked around a ring of finite radius ($\sim 3.4\mu$m) on which myosin motors process with no contractility and/or advection.
{\bf e)}Left Minimal intensity projection (over $120$s) of a spiral-like structure captured using iSCAT (scale bar $2$ $\mu$m) 10-15 min after the addition of 0.1 mM ATP.
{\bf e)}Right Texture and assumed density profile for actin.
\textbf{ f)} For radii $>1\,\mu$m, the myosin density profile broadly captures the angular average of the interferometric scatting intensity.
\textbf{g)} The components of the myosin current have some agreement with density weighted velocities from a PIV analysis, although the oscillatory profiles of the theory are off-set from the experimental data in the direction of the centre of the vortex. 
\textbf{h)} Non-zero, first order corrections to actin velocities due to myosin contractility ensure that the vortex is ultimately unstable on very long timescales.  
Panels {\bf b}-{\bf d} plotted for $\xi=-1$, $\chi_m=1$, $\chi_a=0.1 $, $k=0.8$. Panels {\bf f}-{\bf h} plotted for $\xi=-1$, $\chi_m=1$, $\chi_a=0.1$, $k=10$, and $\kappa=0.1$. Scale bars, $2$$\mu$m.
		}
\end{figure*}

To understand the ramifications of our scaling anzatz, we perform a linear stability analysis of the resulting hydrodynamic theory under perturbations about a uniformly ordered state, deep in the nematic phase ({\it i.e.}, disregarding the lyotropic term in the dynamics of the polarisation vector, and instead imposing $P^i P_i = 1$ via a Lagrange multiplier).

Traditionally, active hydrodynamic descriptions of this type are characterised by an advective-contractile instability, whereby contraction of the actin meshwork advects the bound myosin II motors that are generating the contractile force, resulting in a form of advective feedback that can result in asterlike foci. However, in our case, the ATP dependence of myosin II minifilament off-rates, as well that of the magnitude of active contractility, leads to some modifications. Specifically, each contributes a factor of $k$ ($\propto [ATP]$) to the linear stability condition, resulting in the following quadratic form:
\begin{equation}
  -k-\left(\frac{\xi}{k^2\rho_a^{(0)}}+\chi_m\right)Q^2\geq 0,
  \label{eq:linstab}
\end{equation}
which applies in the small wavevector, $Q$, regime, where $\rho_\text{a}^{(0)}$ is the uniform background actin density (Appendix \ref{sec:app4}). To this order in $Q$, the criterion that there exists an unstable regime is then just given by $k^2 < -\xi/\rho_a^{(0)}\chi_m$, where the right-hand side is positive due to the sign convention that $\xi$ is negative for contractility (and positive for extensility).  That is, advective-contractile feedback is unstable beneath a threshold level of ATP concentration. Of note, by plotting (\ref{eq:linstab}) as a function of $Q$ and $k$ (on log-log scale) alongside a similar condition, but where contractility is ATP-{\it independent}, we can show that, at low enough levels of ATP, a contractile instability can be triggered in a regime where it might otherwise be expected to be stable (Fig.~\ref{fig:numericsFig}b).

Despite this conceptual insight, such analysis is nevertheless performed around a uniformly ordered stable state, rather than the random ordering associated with the deposition of actin filaments in our experiments. We therefore numerically obtained full (nonlinear) solutions to our equations on a periodic domain using a central space, RK4 time finite-difference method (Appendix \ref{sec:app4}). This not only verifies that our ATP-dependent heuristic for advective-contractile feedback holds in the wider setting of our experiments, but also serves to identify ``low'' and``high'' regimes of $k$ for which asters do and do not form, respectively, when starting from uniform densities with random orientation and unit magnitude polarisation. Videos 10 \& 11 show example solutions in high and low regimes, respectively, with corresponding stills from the long time limit shown in Fig.~\ref{fig:numericsFig}b. As expected, in the high ATP case, the actin and myosin densities remain approximately uniform as the polarisation relaxes, and no contractile instability occurs. However, in the low ATP case, an advective-contractile instability occurs and dense actin and myosin clusters form around $+1$ topological defects (aster- or spiral-shaped structures). These structures eventually coarsen to a single aster-like defect where the coarsening time depends on the domain size.

\section{Comparison of steady states between theory and experiment}\label{sec:comparison}

%
Both our analysis and numerical solutions suggest that an ATP-dependent criterion for advective-contractile feedback codifies how actomyosin remodelling changes as ATP depletes. To substantiate this, we take iSCAT data from characteristic vortex (high ATP) and aster (low ATP) textures, and perform angular averages over interferometric contrast and particle image velocimetry (PIV) (Appendix \ref{sec:app4}). We then seek axisymmetric solutions to our equations for each of the two cases, and compare with data using a minimal fitting procedure that excludes points within $1\mu\text{m}$ of the center. This distance is comparable to the size of a single myosin-II minifilament and is therefore the scale at which we expect our continuum theory to break down. 

\subsection{Low ATP Concentration}
At low levels of ATP, $k$ is small, and therefore the magnitude of the active contractile term that appears in (\ref{eq:dimensionlessActinVelcoityExpressions}) via the stress tensor (\ref{eq:sigma}) is large.  In this case, there is an axisymmetric steady-state solution to Eqs.~(\ref{eq:actinDensity}) and (\ref{eq:polarisation}) that is just a stationary radial aster-like actin texture (\textit{i.e.}, $\vec{P}=\vec{e}_r$ and $\vec{v}_a = \vec{0}$).  Since the molecular field is zero in this case, this implies that $\vec{v}_m$ is a function of $\rho_m$, such that the accumulation of myosin II minifilaments due to their procession towards the centre of the texture is balanced by the density-dependent dissociation rate (see Vid.~9 and SM \cite{supp_al-izzi}).  Solving this boundary value problem for $\rho_m$ requires a shooting method.  The solution can be substituted into the remaining force balance condition, which equates contractile forces with those resisting actin compression.  This gives a first-order equation for $\rho_a$ that is easily solvable numerically (Appendix \ref{sec:app4}).

The resulting aster-like steady-state is characterised in Figs.~\ref{fig:hydrodynamicsFig}a-d. In Fig.~\ref{fig:hydrodynamicsFig}a, we show a static iSCAT image along with our assumed aster texture. In Fig.~\ref{fig:hydrodynamicsFig}b, we compare the theoretical myosin density profile, $\rho_{\text{m}}$, with the angular average of the iSCAT scattering intensity, where the overall density scale is fitted by minimising the $L_2$-norm between theory and data (excluding points within $1\mu\text{m}$ of the center). Both the data and theory peak at the centre and then decay and to a constant value in the far field. Similarly, the radial and angular components of the steady-state myosin current can be compared to the density-weighted flow fields inferred from a PIV analysis, Fig.~\ref{fig:hydrodynamicsFig}c. Although the angular average is noisy, this also shows good agreement, with zero angular current (on average) and a decaying radial flow of myosin into the center of the aster. Finally, our steady-state solution predicts that the actin density scales in a similar manner as the myosin density, forming a dense contractile foci at the centre. Whilst this cannot be observed with iSCAT, it compares favourably with earlier, simultaneous fluorescence imaging of myosin and actin in the same \textit{in vitro} system \cite{koster_actomyosin_2016}.

\subsection{High ATP Concentration}
At intermediate to high levels of ATP, $k$ is large, and therefore active contractile stresses are small. Since we further assume that both the elastic modulus $\kappa$ and the inverse compressibility $\chi_m$, are small, this implies that the velocity of actin, $\vec{v}_a$, is also small. We can therefore use a perturbative approach, expanding in $\epsilon=1/k$, and solving hierarchically. We expect the zeroth order solution to be long lived, or quasi-steady, in the sense that any corrections will be $O\left(\epsilon\right)$, and therefore small. 

At zeroth order, $\vec{v}^{(0)}_a = 0$, implying freedom to impose $\vec{P}^{(0)}=\cos\psi^{(0)} \vec{e}_r + \sin\psi^{(0)} \vec{e}_\theta$ and $\rho_a^{(0)}$.  To mimic experimental motifs, we choose a spiral-like texture that transitions between a vortex at large $r$ and an aster as $r\to 0$,
\begin{equation}\label{eq:vortexTexture}
\psi^{(0)} = \frac{\pi}{2}\frac{r^2}{b+r^2},
\end{equation}
with density profile $\rho_a^{(0)} = c r^2 e^{-r^2/w} +l$ (Appendix \ref{sec:app4}).  In this case, the velocity profile of the myosin II minifilaments, $\vec{v}_m$ is such that it balances propulsive forces with density gradients. Substituting into (\ref{eq:myosinDensity}), the resulting boundary value problem is again solvable by a shooting method.

The zeroth order, quasi-steady state solution is shown in Figs.~\ref{fig:hydrodynamicsFig}e-h. Fig.~\ref{fig:hydrodynamicsFig}e shows a minimal intensity projection of the iSCAT data (over $120$s intervals) split with the assumed texture, Eq.~(\ref{eq:vortexTexture}).  Also shown is the initial actin density, whose peak corresponds to circle at $r=3.4\mu\text{m}$. In Fig.~\ref{fig:hydrodynamicsFig}f, we see that, once again, the density of myosin, $\rho_{\text{m}}$, compares well with the iSCAT intensity data for $r>1\mu\text{m}$, although the breakdown within this limit is now much more apparent. The myosin density scale is fitted to the data in the same manner as the aster. 

In order to obtain a comparison between the radial and angular components of the steady-state myosin current and the density-weighted flow fields inferred from a PIV analysis, we must further fix the free parameters of our assumed texture, Eq.~(\ref{eq:vortexTexture}). By minimizing the $L_2$-norm between the quasi steady state currents and the data, we obtain values $(c,l,b)=(0.5,0.1,0.05)$ , see Appendix \ref{sec:app4}. This results in a semi-qualitative fit between our theoretical quasi steady-state solution and the data, Fig.~\ref{fig:hydrodynamicsFig}g. In both cases, there is a small radial current flowing into the centre, and a larger circulating current in the angular direction. However, there is a more pronounced oscillation in the numerical solution than that observed in the data. In the theory, this peaks at exactly the actin ring radius, whereas the data shows a peak closer to the center of the vortex.

There could be several reasons for the discrepancies between our model and the data. The first is that our simple texture(s) may not be representative of the actual actin structure (this is a downside of using iSCAT to visualize the myosin dynamics, as we cannot simultaneously visualize the actin texture). In addition, there may be subtlties associated with the bundling of the long actin filaments used in these experiments (the lengths of which are comparable with the radii of curvature of the vortex structures). Finally, we are likely seeing artefacts of the finite extent of the myosin-II minifilaments. For example, at small scales, a single minifilament may be able to bridge inner sections of the ring-like actin texture.

Ultimately, this quasi steady state is unstable at first order in the actin velocity, suggesting that the ring structure will eventually decay to other structures. We plot the actin flow field for completness in Fig.~\ref{fig:hydrodynamicsFig}g.

\section{Discussion}

When taken together, experiments, single myosin II minifilament kinetics, and bulk hydrodynamics are all consistent with the notion that a change in the concentration of available ATP not only changes the power of actomyosin, but it can also change the mode by which work is done.  Here, we have argued that, as the concentration of available ATP decreases, the processivity of myosin II minifilaments gives way to isotropic contractile forces.  The reason is that the hydrolysis cycle by which chemical energy, in the form of ATP, is transduced into mechanical work involves, as an intetrmediate step, the unbinding of individual heads from actin.  As a result, increased levels of ATP reduce the likelihood that a given myosin II minifilament spans more than one actin filament, therefore reducing the ability for the generation of contractile forces, and instead resulting in an increased likelihood of processive motion.  This heuristic is codified in an ATP-dependent criterion for advective-contractile feedback, the onset of which, as ATP depletes, we propose as the underlying mechanism that dictates a non-trivial relationship between power and remodelling.  

From a theoretical perspective, our work highlights the importance of understanding the functional form of both active and passive coefficients, and the possibility that even simple microscopic models can provide interesting insight and scaling hypotheses. This is relevant because a variety of other active hydrodynamic models, such as those used to study active turbulent-like flows, for example, predict scaling laws that are {\it non}-universal, {\it i.e.,} they depend of the values of the coefficients in theory \cite{martinez-prat_scaling_2021,giomi_geometry_2015,alert_active_2022,dunkel_fluid_2013,thampi_vorticity_2014}.

From a biological perspective, this demonstrates how to encode local biochemistry that, whilst not necessarily hydrodynamic, can nevertheless be crucial to function.
We speculate that our findings may be important for the generic understanding of how actomyosin remodelling is regulated. That is, in addition to the molecular signalling pathways driven by phosphorylation as the canonical mechanism behind the regulation of myosin II activity and/or actin filament growth, ATP concentration may also be important, suggesting a link between essential functions such as contraction or shape change and the metabolic pathways by which ATP is synthesised. Such ideas may even be relevant for understanding the implications of compromised metabolism \cite{jansen_energy_2003,smith_development_2004}.

There are several aspects of our theory and analysis that might be extended. For microscopic modelling, corrections to consider the potential effects of load dependence \cite{debold_slip_2005,erdmann_sensitivity_2016}, cooperativity \cite{walcott_mechanical_2012,kaya_coordinated_2017}, and chirality \cite{tee_cellular_2015}, could lead to quantitative manifestations of such effects at the hydrodynamic level. From a hydrodynamic perspective, the main current challenge is to extend current liquid crystal-like nemato-hydrodynamic descriptions to describe the remodelling of transient networks of intermediate-to-long actin filaments, which are semi-flexible and curved on length scales similar to the filaments themselves.  Whilst such networks have been explored extensively in numerical simulations--- particularly in terms of rigidification transitions and contractility generation \cite{smith_molecular_2007,head_distinct_2003,head_deformation_2003,kasza_actin_2010,stuhrmann_nonequilibrium_2012,tan_self-organized_2018,mulla_origin_2019,alvarado_molecular_2013,koenderink_active_2009,arzash_stress_2019,nedelec_collective_2007}--- writing down a constitutive relation for such materials is a significant outstanding challenge in the field. To our knowledge, it has only been attempted to date at the level of bulk passive rheology and has no straightforward generalisation to active systems \cite{tanaka_viscoelastic_1992-1,tanaka_viscoelastic_1992-2,tanaka_viscoelastic_1992}. Amongst other things, such a constitutive relation would need to account for the known shear-stiffening of actomyosin under imposed strain \cite{koenderink_active_2009}. The Poisson bracket approach for computing reactive couplings in generalised hydrodynamics might prove useful here, as in the nematic polymer theory \cite{chaikin_principles_2000,kamien_poisson_2000}. Here it is worth noting a recent work \cite{redford_motor_2024} which, in similar spirit to our approach, showed how the microscopic ATP dependent kinetics can feed through to bulk hydrodynamic-elastic effects in dense \textit{in vitro} actomyosin.

Regarding our experimental observations, we note that while vortex and spiral formation has been observed in microtubule-based systems \cite{nedelec_self-organisation_1997}, similar works on actomyosin networks did not report such structures \cite{murrell_f-actin_2012, ideses_myosin_2013, alvarado_molecular_2013, koster_actomyosin_2016, smith_molecular_2007, vogel_myosin_2013}. This presents an open question: How precisely do the remarkably persistent vortex- and spiral-like structures that we observe depend on the experimental conditions? Possible factors here include: a relatively low actin filament concentration (equivalent to [G-actin] = 125 nM compared to [G-actin] $> 1 \mu$M in other works) which leads to very thin f-actin layers ($< 6$ filaments) \cite{mosby_myosin_2020}; the absence of crowding factors; the minimisation of laser induced photo-damage by using label-free imaging, and; degassed buffers in combination with an oxygen scavenger system, which decrease the generation of dead heads in myosin II minifilaments that would lead to irreversible crosslinking \cite{alvarado_molecular_2013}.

In conclusion, we believe that our work highlights a novel conceptual advance: ATP is more than `just' a fuel, it impacts the mode by which actomyosin remodels, as well as the rate. In this context, developing more precise experimental setups (\textit{e.g.}, microfluidic control and maintenance of ATP concentrations), more realistic microscopic theories to feed into coarse-grained hydrodynamic parameters, and extending current liquid crystal-like constitutive relations that capture the rheology of transient cross-linked networks are just three of the possible directions future research in this area could take.

\section{Experimental Methods}\label{sec:app1}
\subsection{Purified Proteins}	
Actin was purified from chicken breast following the protocol from Spudich and Watt \cite{spudich_regulation_1971} and kept on ice in monomeric form in G-buffer ($2$ mM Tris Base, $0.2$ mM ATP, $0.5$ mM TCEP-HCl, $0.04$\% NaN3, $0.1$ mM CaCl2, pH $7.0$); for TIRF microscopy experiments, G-actin was labelled with Maleimide-Alexa488 as described earlier \cite{koster_actomyosin_2016}. The experiments using iSCAT microscopy were performed with myosin II obtained from chicken breast following a modified protocol from Pollard \cite{pollard_chapter_1982} and kept in monomeric form in myo buffer ($500$ mM KCl, $1$ mM EDTA, $1$ mM DTT, $10$ mM HEPES, pH $7.0$). For TIRF microscopy experiments, myosin II was purchased (rabbit m.~psoas, Hypermol, \#8306-01). The day before the experiments, functional myosin II proteins were separated from proteins containing dead head domains by a round of binding and unbinding to F-actin in a $5:1$ actin-myosin ratio (switch from no ATP to $3$ mM ATP) followed by a spin at $60,000$ rpm for $10$ min at $4^{\circ}$ C on a TLA100.3 rotor. The supernatant containing functional myosin II was dialysed against myo buffer overnight at $4^{\circ}$ C and could be used for up to three days when stored at $4^{\circ}$C. 

To control the length of actin filaments, we titrated purified murine capping protein to the actin polymerisation mix. The purification of the capping protein was previously described \cite{koster_actomyosin_2016}. To link actin with SLB, we used a construct containing 10x His domains followed by a linker (KCK) and the Ezrin actin binding domain (HKE) as previously described \cite{koster_actomyosin_2016}. For TIRF experiments, we expressed and purified the His10-SNAP-Ezrin actin binding domain (HSE) and labelled it with SNAP surface 549 (New England BioLabs Inc., S91112S) following the manufacturer's protocol. 

\subsection{Supported Lipid Bilayer and Experimental Chamber Preparation}
Experimental chambers were prepared, and supported lipid bilayers (SLB) containing $98\%$ DOPC and $2\%$ DGS-NTA(Ni2+) lipids were formed as previously described \cite{koster_actomyosin_2016} in chambers filled with $100$ $\mu$l KMEH ($50$ mM KCl, $2$ mM MgCl2, $1$ mM EGTA, $20$ mM HEPES, pH $7.2$). The formation of SLB was observed live using iSCAT microscopy, to ensure that a fluid lipid bilayer was formed without holes \cite{mosby_myosin_2020}. 

\subsection{Formation of Acto-Myosin Network}
In a typical experiment, SLBs were formed, incubated with 10 nM HKE for $40$ min, and washed thrice with KMEH. During this incubation time, F-actin was polymerised. First, $10\%$vol of $10\times$ ME buffer ($100$ mM MgCl2, $20$ mM EGTA, pH $7.2$) was mixed with the G-actin stock (for TIRF experiments unlabelled G-actin was supplemented with $10\%$ of Alexa488 labelled G-actin) and incubated for 2 min to replace \(Ca^{2+}\) ions bound to G-actin with \(Mg^{2+}\). The addition of $2\times$ KMEH buffer, supplemented with $2$ mM Mg-ATP, induced the polymerisation of F-actin at a final G-actin concentration of $5$ $\mu$M. After incubation for $20-30$ min, the desired amount of F-actin was added to the SLB using blunt cut $200$ $\mu$ l pipet tips. An incubation of $30$ min allowed the F-actin layer to bind to the SLB at steady state. For the addition of myosin II filaments to the sample, intermediate dilutions of myosin II proteins were prepared in milliQ water at $10$x the final concentration of fresh myosin II stock ($4$mM, $500$mM KCl) in a test tube and incubated for $5$ min to allow the formation of myosin II filaments. Then, $1/10$ of the sample buffer was replaced with the myosin II filament solution and supplemented with Mg-ATP ($100$ mM) at the final concentration of $0.1$ mM. Subsequently, the evolution of the acto-myosin system was observed for up to $60$ min. Typically, the system showed a remodelling behaviour for the first $10-15$ min before contraction and aster formation started (due to ATP concentrations below $10$ $\mu$M as estimated from the activity of myosin II and earlier reports \cite{smith_molecular_2007}). Once the system reached a static jammed state and no myosin activity could be observed, the system could be reset to a remodelling state by adding Mg-ATP ($100$ mM) to a final concentration of $0.1$ mM. Each step of this procedure was performed on the microscope stage, allowing us to continuously check its state. The open chamber design allowed the addition of each component without inducing flows that would perturb the actin network. All experiments were carried out at room temperature ($22^\circ$ C) and buffer evaporation was minimal ($<5\%$vol).
Details about the actin, myosin and ATP concentrations used in each experiment can be found in Table \ref{tab:ExpConditions}.    

\begin{widetext}
\begin{table}[h!]
\centering
 \begin{tabular}{||c || c | c | c | c | c | c||} 
 \hline
 \textbf{Figure} & [Actin]/nM & [CP]/nM & [Myosin II]/nM & [ATP]initial/$\mu$M & anti-bleach & $L_{\text{F-actin}}$/$\mu$m\\
   & & &  &  & agent & \\ \hline
 iSCAT & $300$ & $0.38$ & $100$ & $100$ & & $7\pm 3.5$ \\
experiments  & & & & & & \\ \hline
TIRF & $200$ & $0.25$ & $10$ & $100$ & PCA/PCD & $7.2\pm 1.5$\\
experiments & $200$ & $0.25$ & $40$ & $100$ & PCA/PCD & $7.2\pm 1.5$\\
 & $250$ & $0.25$ & $20$ & $100$ & PCA/PCD & $7.2\pm 1.5$\\
 & $250$ & $0.25$ & $10$ & $100$ & PCA/PCD & $7.2\pm 1.5$\\ \hline
 \end{tabular}
 \caption{Summary of the experimental conditions (final concentrations in the imaging chamber).}
 \label{tab:ExpConditions}
\end{table}
\end{widetext}

\subsection{iSCAT microscope}
Interferometric scattering microscopy was performed on two different custom setups similar to those detailed in \cite{ortega-arroyo_interferometric_2012,cole_label-free_2017}. Briefly, a weakly focused laser beam was scanned across the sample over an area $32.6 \times 32.6$ $\mu\text{m}^2$ ($635$ nm laser). The light reflected from the glass-water interface, together with the scattered light from the sample, was imaged by a CMOS camera (MV-D1024-160-CL-8, Photonfocus, Switzerland). The cameras were controlled using custom LabVIEW software. The videos were recorded at $5$ fps ($635$ nm laser) with the illumination intensity on the sample ($635$ nm laser: $1.9$ $\text{kW cm}^{-2}$) set to nearly saturate the camera with the returning light. The size of the pixels was $31.8$ nm / pixel.

\subsection{Total Internal Reflection Fluorescence (TIRF) microscopy}
TIRF microscopy on acto-myosin networks containing Alexa488 labelled actin and SNAP-surface $549$ labelled HSE were performed on a Nikon Ti Eclipse microscopy equipped with a Nikon TIRF laser coupler, $488$nm, $563$nm and $640$nm lasers, a triple pass TIRF dichroic mirror ($488$nm, $549$nm, $640$nm; Chroma), a $100\times$ ($1.49$NA) Nikon objective and $520$ ($\pm 5$) nm, $615$ ($\pm 5$) nm emission filters (Chroma). Images were recorded with an Zyla sCMOS camera (Andor) controlled by iQ3.0 software (Andor) and the pixel size was $63.4$nm. To limit photodamage, we degassed buffers, employed a radical scavenger system (protocatechuic acid and Protocatechuate 3,4-Dioxygenase, Sigma) and reduced laser intensities to a minimum while imaging the sample only at $5$s per frame.  

\subsection{Image processing (iSCAT)}
Non-sample-specific illumination inhomogeneities, fixed pattern noise, and constant background were removed from the raw images by dividing each of them with a flat field image that contained only these features. The flat-field image was computed by recording 2000 frames of the sample while moving the stage. For each pixel, a temporal median was calculated, resulting in a flat field image that only contained static features.

\subsection{Median filtering}
Movies were median filtered using MATLAB (MathWorks, Natick, MA, USA). For each image sequence, the median is computed for each pixel, deleted from the original image sequence and the median filtered image sequence as well as the computed median filter are saved. 

\subsection{Actin filament length measurements}
Image stacks of actin filaments landing on HKE decorated SLBs were taken at $10$ Hz immediately after adding actin filaments to the sample. These image stacks were split into segments of $10$ s and the median of each segment was subtracted from its last frame to visualise freshly landed isolated actin filaments. The images were then converted from the interferometric contrast values $32$ bit to $8$ bit (using formula $f(x) = -1000 x + 1000$), bandpass filtered (low pass: $3$ pixel, high pass: $20$ pixel) and analysed with the Neurone J image J plugin\cite{meijering_design_2004}. Images of fluorescent actin filaments landing on HSE-decorated SLBs shortly after their addition to the sample were directly converted to $8$ bits and analysed with Neurone J.

\subsection{Particle Image Velocimetry}
Particle Image Velocimetry (PIV) was performed using PIVlab \cite{thielicke_flapping_2014,thielicke_pivlab_2014,garcia_fast_2011}. The median filtered image sequences originating from the iSCAT experiments were inverted as described above and transformed into the $8$-bit format using Image J. The image sequences from the TIRF experiments were filtered by median and directly converted into $8$-bit format using Image-J. The PIV vector maps were computed with the FFT window deformation algorithm and window sizes of $64$, $32$ and $16$ pixels with step sizes of $32$, $16$ and $8$ pixels for iSCAT experiments, and window sizes of $32$ and $16$ pixels with step sizes of $16$ and $8$ pixels for TIRF experiments, respectively. The accepted range of velocity vectors was limited to $\pm 2$ $\mu\text{m s}^{-1}$. 

\subsection{ATP Concentration Assays}
\subsubsection{Myosin II Preparation}

The day prior to the primary experiment, Hypermole myosin II was diluted to one-fourth of its original concentration, resulting in a final volume of $ 400 \,\mu L$ with a concentration of $0.5 \, \mu M$. This diluted myosin was dialyzed against a dialysis buffer composed of 500 mM KCl, 10 mM HEPES, 1 mM EGTA, and 1 mM $MgCl_2$.
\subsubsection{Supported Lipid Bilayer (SLB) Preparation and Ezrin Incubation}
On the day of the experiment, 5 to 7 SLBs were prepared and incubated with 10 nM Ezrin for 40 minutes. Following incubation, the SLBs were washed three times with KMEH buffer. Concurrently, F-actin was polymerized. After 20 minutes of incubation, F-actin was added to the SLBs using cut pipette tips to achieve a final concentration of 300 nM, followed by three washes with KMEH buffer.
\subsubsection{Myosin Supplementation and Sample Preparation}
The dialysed myosin, at a concentration of 0.5 µM, was supplemented with 0.5 mM ATP.ATP-supplemented myosin was added to the sample buffers, replacing one fifth of the buffer to reach a final ATP concentration of 0.1 mM. At various time points (e.g. 0.5, 5, 13, 16, and 20 minutes), 50 µL samples were extracted from different SLBs. These samples were centrifuged at 60,000 rpm for 20 minutes at \(4^{\circ}\) to remove excess actin and myofilaments. Subsequently, 45 µL of the supernatant was collected.

\subsubsection{ATP Assay}
To determine the ATP concentrations of our samples, we used a colorimetric ATP assay (PhosphoWorks, cat\# 21617, AAT bioquest)  according to the manufacturer's manual. The collected samples were transferred to a 96-well plate (flat bottom), to which 45 µL of ATP assay working solution was added. Control samples were prepared in conjunction with the experimental samples, including ATP standards for calibration (at $100\mu\text{M}$, $50\mu\text{M}$, $25\mu\text{M}$ and $12.5\mu\text{M}$ ATP), SLBs containing only actin without ATP or myosin, and SLBs containing only myosin without actin or ATP. KMEH buffer was used as a blank. The plate was sealed with foil and incubated for $35-45$ minutes.  Absorbance at $570$ nm was recorded for each sample using a microplate reader (Varioskan, Thermofisher), and background absorbance from the blank control was subtracted from all readings. A linear calibration graph of absorbance versus ATP concentration was created from the data of the standard samples. This linear calibration function was used to convert readings from the collected samples into ATP concentrations.

\begin{figure}
    \centering
    \includegraphics[width=0.48\textwidth]{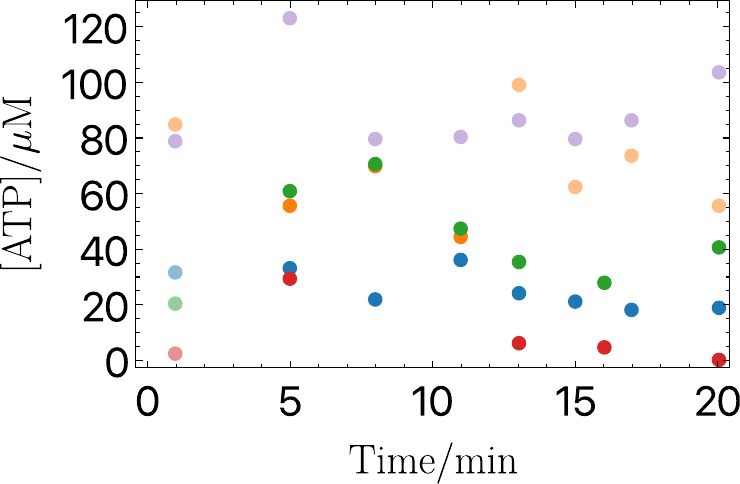}
    \caption{Full data from ATP concentration assays as a function of time. The first time-point is neglected as the assay consistently under-reports (the know concentration at $t=0$ is $100\mu$M). If any later data point reports above $95\mu$M we exclude it and all further reads from that experiment. The excluded points are shown with half opacity, with purple a control experiment in the absence on myosin motors.}
    \label{fig:fullATPData}
\end{figure}


\acknowledgements{RGM and SCA-I acknowledge funding from the EMBL-Australia program and the Australian Research Council Centre of Excellence for the Mathematical Modelling of Cellular Systems (MACSYS, CE230100001). DVK thanks Phillip Kukura and Nicolas Hundt for enabling the iSCAT experiments, Mohan Balasubramanian for access to the TIRF microscope, and the Wellcome-Warwick Quantitative Biomedicine Programme  (Wellcome Trust ISSF) and EPSRC (EP/V043498/1) for funding. SGN was supported by the EPSRC (EP/V043498/1). SCA-I would like to thank Jack Binysh \& Richard Spinney for helpful comments and suggestions.}

\makeatletter 
\def\tagform@#1{\maketag@@@{(\ignorespaces#1\unskip\@@italiccorr)}}
\makeatother
\graphicspath{{figures/}} 
\def\eq#1{{\ref{#1}}}    
\def\fig#1{{\ref{#1}}}
\appendix

\section{Microscopic Model}\label{sec:microscopicModels}\label{sec:app3}
Here, we include the details of the calculations for our microscopic model for the statistics of the bound status of the individual head groups beloonging to a single myosin II minifilament (see Fig.~2 of main text).

\subsection{Master Equations}
In our model, the time-dependent probabilities $P_n(t)$ for a single myosin II minifilament to have $n$ heads bound are given by the following coupled, linear ODEs:
\begin{align}\label{eq:filamentMasterEqs}
&\frac{\mathrm{d}P_0(t)}{\mathrm{d}t} = k_{\text{off}}P_{1}(t),\\
&\frac{\mathrm{d}P_1(t)}{\mathrm{d}t} = 2k_{\text{off}}P_2(t)-(N-1)k_{\text{on}}P_1(t)-k_{\text{off}}P_1(t),\\
&\frac{\mathrm{d}P_n(t)}{\mathrm{d}t} = (N-n+1)k_{\text{on}}P_{n-1}(t)+(n+1)k_{\text{off}}P_{n+1}(t)\nonumber\\
& \quad -\left[(N-n)k_{\text{on}}+n k_{\text{off}}\right]P_n(t),\quad \forall\ n\in [2,N-1],\\
&\frac{\mathrm{d}P_N(t)}{\mathrm{d}t} = k_{\text{on}}P_{N-1}(t)-N k_{\text{off}}P_{N}(t),\label{eq:filamentMasterEqsEnd}
\end{align}
with initial condition $P_{1}(0)=1$ (\textit{i.e.}, one head bound) representing an the first state of the minifilament on binding from the bulk. We are interested in the average dwell time of the myosin filament on the actin filament, or the mean first passage time into the $n=0$ state (\textit{i.e.}~dissociation from actin back to the bulk).

\subsection{Recursive solution for mean first passage time starting from one head bound}
The distribution of such first passage times to dissociation is given by
\begin{equation}
f_0(T)=\text{Pr}\left\{x(T)=0,\ x(t)\neq0\ \forall\ 0\leq t<T\ |\ x(0)=1\right\},
\end{equation}
with mean
\begin{equation}
\langle T_{0}\rangle = \int_0^\infty t f_0(t)\mathrm{d}t = - \partial_s\left[\bar{f}_{0}(s)\right]|_{s=0},\label{eq:T_0}
\end{equation}
where $\bar{f}_0(s)$ denotes the Laplace transform of $f_0(t)$. Taking the Laplace transform of the master equations (\eq{eq:filamentMasterEqsEnd}) gives
\begin{align}
&s\bar{P}_0= k_{\text{off}}\bar{P}_{1},\\
&s\bar{P}_1-1 = 2k_{\text{off}}\bar{P}_2-(N-1)k_{\text{on}}\bar{P}_1-k_{\text{off}}\bar{P}_1,\\
&s\bar{P}_n = (N-n+1)k_{\text{on}}\bar{P}_{n-1}+(n+1)k_{\text{off}}\bar{P}_{n+1}\nonumber\\
&\quad-\left[(N-n)k_{\text{on}}+n k_{\text{off}}\right]\bar{P}_n,\quad \forall\ n\in [2,N-1],\\
&s\bar{P}_N = k_{\text{on}}\bar{P}_{N-1}-N k_{\text{off}}\bar{P}_{N},
\end{align}
which can be solved recursively to give
\begin{align}
&\frac{\bar{P}_n}{\bar{P}_{n-1}} = \frac{\left(N-n+1\right)k_{\text{on}}}{s+\left(N-n\right)k_{\text{on}}+ n k_{\text{off}} - \left(n+1\right)k_{\text{off}}\frac{\bar{P}_{n+1}}{\bar{P}_{n}}},\nonumber\\
& \qquad\forall\ n\in [2,N],\\
&\bar{P}_1 = \frac{1}{s+\left(N-1\right)k_{\text{on}}+ k_{\text{off}} - 2 k_{\text{off}}\frac{\bar{P}_{2}}{\bar{P}_{1}}},\\
&s \bar{P}_0 = k_{\text{off}}\bar{P}_1 = \frac{k_{\text{off}}}{s+\left(N-1\right)k_{\text{on}}+ k_{\text{off}} - 2 k_{\text{off}}\frac{\bar{P}_{2}}{\bar{P}_{1}}},
\end{align}
with termination condition $\bar{P}_{N+1}=0$. 

The mean first passage, or equivalently, mean residence times, $\langle T_0\rangle$, for $N\in [1,5]$ are plotted in Fig.~\ref{fig:microscopicModel}c as a function of the ratio of off- to on-rates, $k_{\text{off}}/k_{\text{on}}$. In the limit of large $k_{\text{on}}$, the dwell time scales like $\langle T_0\rangle\sim k_{\text{on}}^{N-1}$, whereas for large $k_{\text{off}}$, we have $\langle T_0\rangle\sim k_{\text{off}}^{-1}$.

The functional form of the dwell times for $N\in [1,5]$ are given by:
\begin{align*}
&N=1:\quad \langle T_0\rangle=\frac{1}{k_{\text{off}}},\\
&N=2:\quad \langle T_0\rangle=\frac{2 k_{\text{off}}+k_{\text{on}}}{2 k_{\text{off}}^2},\\
&N=3:\quad \langle T_0\rangle=\frac{3 k_{\text{off}}^2+3 k_{\text{off}} k_{\text{on}}+k_{\text{on}}^2}{3 k_{\text{off}}^3},\\
&N=4:\quad \langle T_0\rangle=\frac{(2 k_{\text{off}}+k_{\text{on}}) \left(2 k_{\text{off}}^2+2 k_{\text{off}} k_{\text{on}} + k_{\text{on}}^2\right)}{4 k_{\text{off}}^4},\\
&N=5:\quad \nonumber\\
&\langle T_0\rangle=\frac{5 k_{\text{off}}^4+10 k_{\text{off}}^3 k_{\text{on}}+10 k_{\text{off}}^2 k_{\text{on}}^2+5 k_{\text{off}} k_{\text{on}}^3 + k_{\text{on}}^4}{5 k_{\text{off}}^5}\text{.}
\end{align*}

\subsection{Sum-over paths representation}
Here we outline a sum-over-paths approach, which is an alternate method of calculating the mean first passage time that also permits the computation of the mean number of visits to each state in our microscopic model \cite{budnar_anillin_2019}.

Consider a sequence of jumps on the ladder state-space that represents number of bound heads of a single myosin II minifilament (see Fig.~\ref{fig:microscopicModel}a ) beginning with one head attached and terminating when the filament drops off (\textit{i.e.}, 0 heads bound). The probability of realising this sequence of jumps is given by the sum over the probabilities of all such trajectories
\begin{widetext}
\begin{equation}
\mathrm{Pr}\left(\mathcal{S}=(1,\dots,0)\right) = \int_{t_0<\dots<t_n}\left(\prod_{i=0}^{n}\mathrm{d}t_i\right) \mathrm{Pr}\left(\mathcal{T}=((1,t_0),\dots,(0,t_n))\right),
\end{equation}
\end{widetext}
where the probability of a given trajectory is just given by the product of the transition probabilities making up that trajectory
\begin{equation}
 \mathrm{Pr}\left(\mathcal{T}=((1,t_0),\dots,(0,t_n))\right) = \prod_{i=0}^{n-1}Q_{i\to j}(t_{i+1}-t_i).
\end{equation}
The transition probabilities are themselves given by the product of the waiting time distributions, $P_{i\to j}(t)$, and survivor functions, $S_{i\to k}(t)$:
\begin{equation}
Q_{i\to j}(t) = P_{i\to j }(t)\prod_{k\neq i,j}S_{i\to k}(t),
\end{equation}
where $S_{\i\to k}(t)=1-\int_0^{t}P_{i\to k}(\tau)\mathrm{d}\tau$. In general, the waiting time distributions are exponential (\textit{i.e.}, Poissonian) with a form $P_{i\to j}(t) = \lambda_{i\to j}\exp\left(-\lambda_{i\to j}t\right)$, where the rates $\lambda_{i\to j}$ can be read off the master equation, Eqs.~(\eq{eq:filamentMasterEqs}--\eq{eq:filamentMasterEqsEnd}). By making use of the convolution properties of the Laplace transform we see 
\begin{equation}
\mathrm{Pr}\left(\mathcal{S}=(1,\dots,0)\right) = \prod_{i=0}^{n-1}\bar{Q}_{i\to j}(0),
\end{equation}
where the bar denotes the Laplace transform. From this, we define a matrix $\mathsf{A}$ with coefficients,
\begin{equation}
A_{ij}=\bar{Q}_{i\to j}(s).
\end{equation}
Raising $\mathsf{A}$ to the power $n$ and taking the $i0$ component (\textit{i.e.}, $[\mathsf{A}^n]_{i0}$) is then just the Laplace transform of the distribution of first passage times into the absorbing state, $0$, in $n$ steps, starting from initial state $i$. Thus, taking $i=1$, summing over all possible paths and taking the Neumann sum we find
\begin{equation}
\bar{f}_{0}(s) = \left[\sum_{n=0}^{\infty} \mathsf{A}^n\right]_{i0} = \left[\left(\mathbb{I}-\mathsf{A}\right)^{-1}\right]_{i0}\text{,}
\end{equation}
from which $\langle T_0\rangle$ can be readily calculated [see Eq.~(\eq{eq:T_0})]. To calculate the average number of times a particular transition happens, $\#\left(i\to j\right)$, we define the matrix $\mathsf{B}$, with coefficients
\begin{equation}
B_{kl}(z) = \begin{cases}
z\,A_{kl}(0)\quad \forall\ (k,l)=(i,j)\\
A_{kl}(0)\quad \text{otherwise}
\end{cases},
\end{equation}
so that the corresponding Neumann series gives the generating function of the probability distribution of a sequence where $\#\left(i\to j\right)=m$---  \textit{i.e.},~some sequence, starting with 1 head bound and ending in dissociation, where the transition $i\to j$ occurs $m$ times:
\begin{equation}
G(z) = \sum_{m=0}^{\infty}z^m P(m) = \left[\sum_{n=0}^{\infty} \mathsf{B}^n\right]_{10} = \left[\left(\mathbb{I}-\mathsf{B}\right)^{-1}\right]_{10}.
\end{equation}
As a result, it follows that
\begin{equation}
\langle\#\left(i\to j\right)\rangle=\langle m \rangle = \partial_zG(z)|_{z\to 1}.
\end{equation}
To illustrate via a concrete example, consider the case of 5 heads, for which we have
\begin{widetext}
\begin{equation}
\mathsf{A}(s) = \left(
\begin{array}{cccccc}
 0 & 0 & 0 & 0 & 0 & 0 \\
 \frac{k_{\text{off}}}{k_{\text{off}}+4 k_{\text{on}}+s} & 0 & \frac{4 k_{\text{on}}}{k_{\text{off}}+4 k_{\text{on}}+s} & 0 & 0 & 0 \\
 0 & \frac{2 k_{\text{off}}}{2 k_{\text{off}}+3 k_{\text{on}}+s} & 0 & \frac{3 k_{\text{on}}}{2 k_{\text{off}}+3 k_{\text{on}}+s} & 0 & 0 \\
 0 & 0 & \frac{3 k_{\text{off}}}{3 k_{\text{off}}+2 k_{\text{on}}+s} & 0 & \frac{2 k_{\text{on}}}{3 k_{\text{off}}+2 k_{\text{on}}+s} & 0 \\
 0 & 0 & 0 & \frac{4 k_{\text{off}}}{4 k_{\text{off}}+k_{\text{on}}+s} & 0 & \frac{k_{\text{on}}}{4 k_{\text{off}}+k_{\text{on}}+s} \\
 0 & 0 & 0 & 0 & \frac{5 k_{\text{off}}}{5 k_{\text{off}}+s} & 0 \\
\end{array}
\right),
\end{equation}
\end{widetext}
which implies a mean first passage time to dissociation of the form
\begin{equation}
    \langle T_0\rangle = \frac{5 k_{\text{off}}^4+10 k_{\text{off}}^3 k_{\text{on}}+10 k_{\text{off}}^2 k_{\text{on}}^2+5 k_{\text{off}} k_{\text{on}}^3+k_\text{on}^4}{5 k_\text{off}^5}\text{.}
\end{equation}

Using the same example, but picking out the transition $1\to 2$, we write 
\begin{widetext}
\begin{equation}
\mathsf{B}(s)=\left(
\begin{array}{cccccc}
 0 & 0 & 0 & 0 & 0 & 0 \\
 \frac{k_{\text{off}}}{k_{\text{off}}+4 k_{\text{on}}+s} & 0 & \frac{4 k_{\text{on}} z }{k_{\text{off}}+4 k_{\text{on}}+s} & 0 & 0 & 0 \\
 0 & \frac{2 k_{\text{off}} }{2 k_{\text{off}}+3 k_{\text{on}}+s} & 0 & \frac{3 k_{\text{on}}}{2 k_{\text{off}}+3 k_{\text{on}}+s} & 0 & 0 \\
 0 & 0 & \frac{3 k_{\text{off}}}{3 k_{\text{off}}+2 k_{\text{on}}+s} & 0 & \frac{2 k_{\text{on}}}{3 k_{\text{off}}+2 k_{\text{on}}+s} & 0 \\
 0 & 0 & 0 & \frac{4 k_{\text{off}}}{4 k_{\text{off}}+k_{\text{on}}+s} & 0 & \frac{k_{\text{on}}}{4 k_{\text{off}}+k_{\text{on}}+s} \\
 0 & 0 & 0 & 0 & \frac{5 k_{\text{off}}}{5 k_{\text{off}}+s} & 0 \\
\end{array}
\right),
\end{equation}
\end{widetext}
and therefore
\begin{equation}
    \langle \# 1\to 2\rangle = \frac{4 k_\text{on}}{k_\text{off}}\text{.}
\end{equation}

\begin{figure}
\center\includegraphics[width=0.4\textwidth]{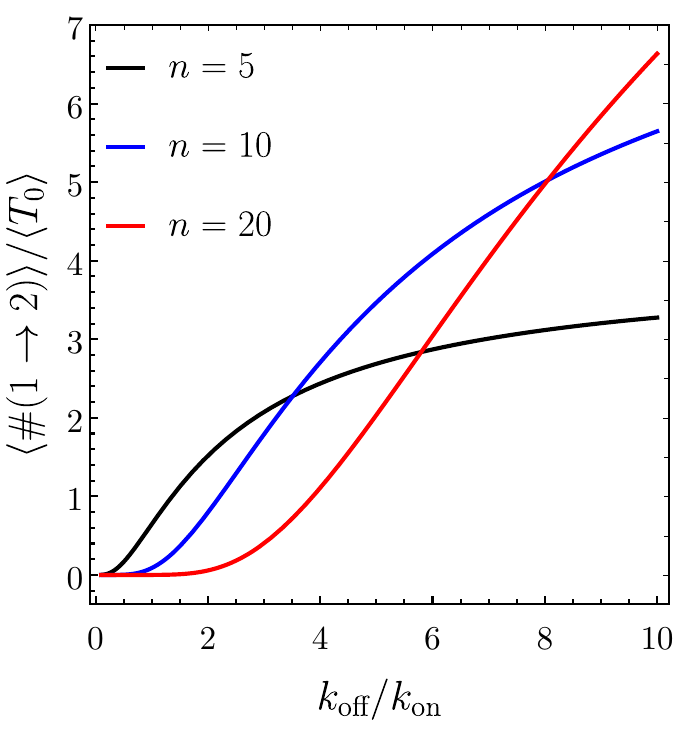}
\caption{\label{fig:powerStrokes}Average number of times a transition from two heads to one head bound occurs, divided by mean first passage time (\textit{i.e.}, $\langle\#\left(1\to 2\right)\rangle/\langle T_0\rangle$) plotted against $k_{\text{off}}/k_{\text{on}}$ for different total number of heads on the myosin II minifilament, $n$. We identify this transition with a processive-like power stroke due to the lack of crosslinking.}
\end{figure}

\begin{figure*}
\center\includegraphics[width=0.8\textwidth]{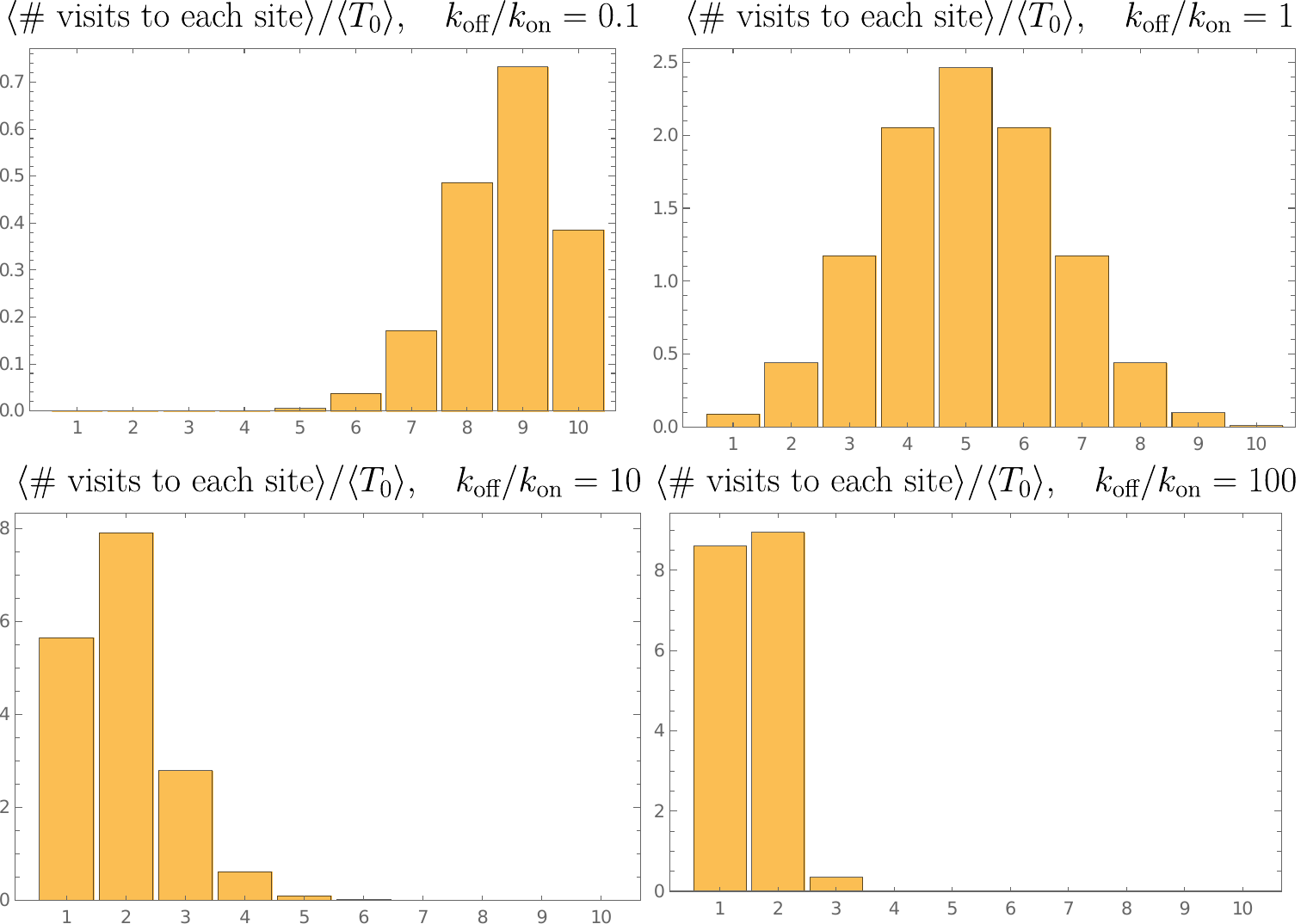}
\caption{\label{fig:numberVisitsCombined}Average number of visits per site per unit dwell time for a filament with $10$ heads plotted for different values of $k_{\text{off}}/k_{\text{on}}$.}
\label{fig:hist}
\end{figure*}

Under the assumption that every transition between states of increasing number--- 
\textit{i.e.}, $i\to i+1$ \textit{etc}.--- involves a single ``powerstroke'', we can further categorise such transitions and therefore their rates.  For example, we identify the transition $1\to 2$ with processive motion; when myosin II minifilaments take a step along actin.  In this case, the ratio $\langle\#\left(1\to 2\right)\rangle/\langle T_0\rangle$ quantifies the rate at which a single minifilament makes processive steps on average. This is plotted for differing numbers of heads and $k_{\text{off}}/k_{\text{on}}$ in Fig.~\fig{fig:powerStrokes}. A histogram of such average rates, over all transitions, can also be plotted (see Fig.~\fig{fig:hist} for $N=10$).  This shows that, as the off-rate is increased, the filament spends more of its time in the states with only a few heads bound and therefore has an increased likelihood of processive motion (note, uninterrupted walking corresponds to the sequence: $1\to 2\to 1\to 2\ldots$ \textit{etc}.).

If \textit{more} than 2 head groups are bound, we assume one of two scenarios.  Either, the additional bindings stall any attempted procession along actin, or, the head groups are bound across more than one actin filament, and the powerstroke results in relative motion of two or more filaments, \textit{i.e.} contractility. Summing the number of ``up'' transitions per mean first passage time for each regime gives a processivity and contractility respectively which are plotted in Fig.~\ref{fig:microscopicModel}d. For large $k_{\text{off}}/k_{\text{on}}$ the processive rate is approximately constant and the contractility scales like $k_{\text{on}}/k_{\text{off}}$, we will make use of this crude scaling in our hydrodynamic model and use it to gain some insight into the ATP dependence of structures seen in experiment.

\section{Hydrodynamic Model}\label{sec:app4}
Here we present the details and solutions to the hydrodynamic model presented in the main text. We take a typical active hydrodynamics approach for a polar gel.  This comprises conservation equations for both myosin and actin densities, force balance conditions that apply between the SLB and actin, and actin and myosin, and a dynamical equation for the broken symmetry field representing the local average of the direction (towards the plus end) of the the actin at each point \cite{marchetti_hydrodynamics_2013,husain_emergent_2017,gowrishankar_nonequilibrium_2016,kruse_generic_2005,prost_active_2015,kruse_asters_2004}.

\subsection{Derivation of dynamical equations}
We assume that the system can be modelled as a triple layered system of myosin-II minifilaments atop actin, which in turn is crosslinked \textit{via} ezrin to a supported lipid bilayer (SLB). We will treat the SLB as a momentum sink such that it only enters the equations for actin via a friction term and does not require the solving of the $2$D Stokes (or D'Arcy equations) in the presence of point forces representing ezrin.

The actin density, $\rho_a$, and myosin density, $\rho_m$, obey continuity equations with source and sink terms for the latter representing the binding/unbinding of minifilaments from the bulk. These are given by
\begin{align}
&\partial_t\rho_a + \nabla_iJ^i_a=0\text{,}\\
&\partial_t\rho_m +\nabla_iJ^i_m =k^m_{\text{on}}\frac{\rho_a}{\rho_a+\rho_a^h}-k^m_{\text{off}}\,\rho_m/\phi\text{,}
\end{align} 
where we have assumed a saturating Hill-form (with respect to actin density) for the myosin minifilament on-rate  as well as an off-rate that is linear in myosin minifilament density. The actin and myosin currents are given by $J^i_{m,a} = v^i_{m,a}\rho_{m,a}$ respectively, where $v^i_{m,a}$ are the components of myosin and actin velocities.

Since the actin is characterised by a polarization (broken symmetry) field, $\vec{P}$, the total stress in the actin layer is given by the tensor $\Sigma^{ij}=\sigma^{ij} + \frac{1}{2}\left(P^ih^j-h^iP^j\right)$, where $h^i$ is the molecular field. The molecular field is derived from from an elastic Frank free energy that we assume to be in the one-constant approximation:
\begin{equation}
\mathcal{F}=\int \frac{\kappa}{2}|\nabla_i P^j|^2\mathrm{d}^2x; \quad h^i=-\frac{\delta \mathcal{F}}{\delta P_i}= \kappa \Delta P^i
\end{equation}
where $\kappa$ is the elastic modulus and $\Delta=\nabla_i\nabla^i$ is the Laplacian. In low density suspensions with long filaments the there is often reason to assume that the bend and splay moduli will differ significantly, and that the bend elasticity will be much larger than the splay. Our experiments suggest that there is minimal passive reordering of the actin filaments and the molecular field has a minimal effect on the dynamics. We performed the simulations both within the one-constant approximation and with a purely bend elastic free energy, due to the small value of the elastic constants we discerned no notable differences in the dynamics or steady states (with the exceptions of spiral-like textures being marginally suppressed in the low ATP steady state for the purely bend free energy). A way to intuitively see this is that in the quasi steady states the elastic terms enter only in the higher order correction to the actin velocity and for the full steady state the aster is an exact solution for both the one-constant and pure bend model. Because of this we elected to show only the results from the one-constant approximation for simplicity, as this is significantly less complex analytically.

We assume a fluid like constitutive relation for the actin in the frictional limit, where the dominant dissipative term is friction with the membrane rather than intra-actin viscosity. The symmetric part of the actin stress tensor is therefore given by the following relation for an active ordered fluid
\begin{equation}
\sigma^{ij} = -\zeta(\vec{r}) g^{ij} - \bar{\zeta}(r) P^iP^j + \left(\chi_m\rho_m-\chi_a\rho_a\right)g^{ij},
\end{equation}
where $\zeta$ and $\bar{\zeta}$ are the active isotropic and anisotropic stresses, and $g^{ij}$ is the inverse of the metric. Assuming that the friction between the myosin and actin layer is given by $\Gamma_m=\beta_m\rho_m$ and between the actin and SLB by $\Gamma_a=\beta_a\rho_a$, the corresponding force balance conditions are then given by
\begin{equation}
\Gamma_m\left(v^i_m-v^i_a\right)= - \alpha v_0\beta_m\rho_m P^i - \chi_m\nabla^i\rho_m,
\label{eq:force_balance_myosin}
\end{equation}
and
\begin{equation}
\Gamma_a v^i_a + \Gamma_m\left(v^i_a - v^i_m\right)= \alpha v_0\beta_m\rho_m P^i + \nabla_j\Sigma^{ij},
\label{eq:force_balance_actin}
\end{equation}
respectively, where $v_0$ is a characteristic rate of myosin procession along actin, and $\alpha$ is a scalar that encodes the dependence of this velocity on non-hydrodynamic variables such as the concentration of ATP.

The objective rate for the polarization, $P^i$, is given by $\mathrm{D}_t P^i=\partial_t P^i + v^j_a\nabla_j P^i + \Omega^{i}{}_{j} P^j$, where $\Omega_{ij}=\frac{1}{2}\left(\nabla_{i}v_{a}{}_{j}-\nabla_j v_{a}{}_{i}\right)$ is the vorticity. This appears in the dynamical equation for the polarization field, which is given by
\begin{equation}
\mathrm{D}_{t} P^i = \frac{h^i}{\gamma} + \nu\left[\frac{\rho_a}{\rho_a^\star}(1-P_jP^j) -(1+P_jP^j) \right] P^i\text{,}
\end{equation}
where the second term ensures $|P|\to 1$ when $\rho_a>\rho_a^\star$.  When deep in the nematic phase, we can approximate this as
\begin{equation}
\mathrm{D}_{t} P^i = \frac{h^i}{\gamma} +\lambda P^i\text{,}
\end{equation}
where we introduce a Lagrange multiplier, $\lambda$, in order to impose $|\vec{P}|=1$. Here we will also assume the actin is sufficiently dense that the on-rate of myosin motors saturates. This gives a simplified continuity equation of the following form
\begin{equation}
    \partial_t\rho_m +\nabla_iJ^i_m = k^m_{\text{on}} - k^m_{\text{off}}\rho_m/\phi\text{.}
\end{equation}

\subsection{Steady-state equations in the axisymmetric case}
We will consider a simplified scenario for steady states where all variables are axisymmetric such that the depend only on the radial coordinate, $r$, and not on the polar angle, $\theta$. We assume that we are in the dense, nematic phase and look for solutions in steady state.

The polarization field we choose to be given by $\vec{P}=\cos\psi(r)\vec{e}_r +\sin\psi(r)\vec{e}_\theta$ and we decompose the molecular field into parallel and perpendicular components 
\begin{align}
&h_{\parallel}=h_r\cos\psi +h_\theta\sin\psi\\
&h_{\perp}=-h_r\sin\psi +h_\theta\cos\psi
\end{align}
where $h_r$ and $h_\theta$ are the components of $\vec{h}$ in polar coordinates. From the one-contant Frank free energy we find
\begin{equation}
    h_{\perp} = \kappa \left[\partial_r^2\psi +\frac{\partial_r\psi}{r}\right]\text{.}
\end{equation}

The continuity equations simplify to
\begin{align}
&k_{\text{off}}\rho_m/\bar\phi-k_{\text{on}}+ \rho_m \partial_r v_m^{r}+ v_m^{r} \partial_r\rho_m+ \frac{\rho_m  v_m^{r}}{r}=0\\
&\rho_a \partial_rv_a^r+ v_a^r \partial_r\rho_a + \frac{\rho_a v_a^r}{r}=0\text{.}
\end{align}

The components of the actin stress tensor, $\sigma$, are given by the following
\begin{align}
&rr:\quad-\frac{1}{2} \left\{\bar{\zeta}\cos (2 \psi)+\bar{\zeta} +2 \zeta\right\}\\
&\theta\theta:\quad \frac{1}{2} \left\{\bar{\zeta}\cos (2 \psi)-\bar{\zeta} -2 \zeta\right\}\\
&r\theta:\quad -\frac{1}{2} \bar{\zeta}\sin (2 \psi)\text{.}
\end{align}

The polarization equation can be re-written as
\begin{align}
&\frac{h_\parallel}{\gamma }=0\\
&-\frac{h_\perp}{\gamma }- \frac{1}{2}\partial_rv_a^\theta +\frac{v_a^\theta +2 r v_a^r \partial_r\psi}{2r} = 0
\end{align}

The components of the myosin force balance equation are given by
\begin{align}
&\vec{e}_r:\quad  \Gamma_m\left(v_m^r- v_a^r\right)=-\alpha  \cos (\psi)- \chi_m\partial_r\rho_m \\
&\vec{e}_\theta:\quad \Gamma_m\left(v_m^\theta-v_a^\theta\right)=-\alpha \sin (\psi)
\end{align}
where $\alpha= v_0\beta_m\rho_m$ and $\Gamma_m = \beta_m\rho_m$.

This tensor divergence has the following components in polar coordinates
\begin{align}
&\vec{e}_r:\quad \partial_r\Sigma_{rr} +\frac{1}{r}\partial_\theta\Sigma_{r\theta} +\frac{\Sigma_{rr}-\Sigma_{\theta\theta}}{r}\\
&\vec{e}_\theta:\quad \partial_r\Sigma_{\theta r} +\frac{1}{r}\partial_\theta\Sigma_{\theta \theta} +\frac{\Sigma_{\theta r}+\Sigma_{r\theta}}{r}
\end{align}
and hence the actin force balance by
\begin{widetext}
\begin{align}
&\vec{e}_r:\quad \frac{1}{2r}\Bigg\{\cos (2 \psi) \left[r \left(  \partial_r\bar{\zeta}\right) + 2 \bar{\zeta}\right] +r \left( \partial_r\bar{\zeta} +2 \partial_r\zeta + 2 \partial_r\Pi_a\right)-\sin (2 \psi) \left(2 r \partial_r\psi \bar{\zeta}\right)\Bigg\}\nonumber\\
& - \Phi \cos (\psi )+ \Gamma_a v_a^r + \Gamma_m\left(v_a^r-v_m^r\right)=0\\
&\vec{e}_\theta:\quad \frac{1}{2r}\Bigg\{\sin (2 \psi) \left[r  \partial_r\bar{\zeta} + 2 \bar{\zeta}\right] +\cos (2 \psi) \left(2 r \partial_r\psi \bar{\zeta}\right)-r \partial_r h_\perp\Bigg\} \nonumber\\
&-\Phi \sin (\psi)+ \Gamma_a v_a^\theta + \Gamma_m\left(v_a^\theta-v_m^\theta\right)=0
\end{align}
\end{widetext}
where $\Pi_a= \chi_m\rho_m -\chi_a\rho_a$, $\Phi= \alpha v_0\beta_m\rho_m$, $\Gamma_a = \beta_a\rho_a$ and $\Gamma_m = \beta_m\rho_m$.

\subsection{Non-dimensionalisation and scaling of the active stress}
Taking the myosin continuity equation, we can write it in dimensionless form as
\begin{equation}
\partial_t\rho_m +\vec{\nabla}\cdot\left(\rho_m\vec{v}_m\right)=\frac{k_{\text{on}}^{(0)}\tau}{\phi}\left[1-k\rho_m\right]
\end{equation}
where $\tau$ is some characteristic time-scale and $k=k_{\text{off}}/k_{\text{on}}$. We will choose $\tau=\phi/k_{\text{on}}$ as our characteristic time-scale as we expect the characteristic time to scale like the dwell time which goes like the inverse of $k_{\text{off}}$ as ATP is varied, see Sec.~\ref{sec:microscopicModels}. We have assumed $\phi$ is the characteristic density scale. The characteristic time-scales and length-scales are given by
\begin{equation}
\tau=\frac{\phi}{k_{\text{on}}}; \quad L=v_0\tau
\end{equation}
where $v_0$ is the active precession velocity of myosin. The dimensionless equation for myosin density then becomes
\begin{equation}
\partial_t\rho_m +\vec{\nabla}\cdot\left(\rho_m\vec{v}_m\right)=1-k\rho_m\text{.}
\end{equation}

From this we can proceed to non-dimensionalise the equations of motion where now all the coefficients are dimensionless as follows $\chi_m\phi/(L \beta_m)\to \chi_m$, $\chi_a\phi/(L \beta_a)\to \chi_a$, $\xi\tau/\beta_a\phi L\to \xi$ and $\tau/\gamma\to 1/\gamma$. This gives the equations quoted in the main text, Eqs.~(1--6). We will also assume that $\chi_a\sim 1/k$ in terms of magnitude.

The active parameter in our model have been chosen to be of the following form
\begin{equation}
\zeta = \frac{ \rho_m}{k} \xi; \quad \bar\zeta = \frac{ \rho_m}{k} \bar\xi
\end{equation}
where we will label the prefactor in the high ATP regime (large $k$) as $1/k=\epsilon$ . This form gives the correct scaling in terms of $k_{\text{off}}$ in the limit of low myosin density.

\subsection{Neglecting anisotropic active stress}\label{sec:anisoTerms}
\begin{figure}
\includegraphics[width=0.4\textwidth]{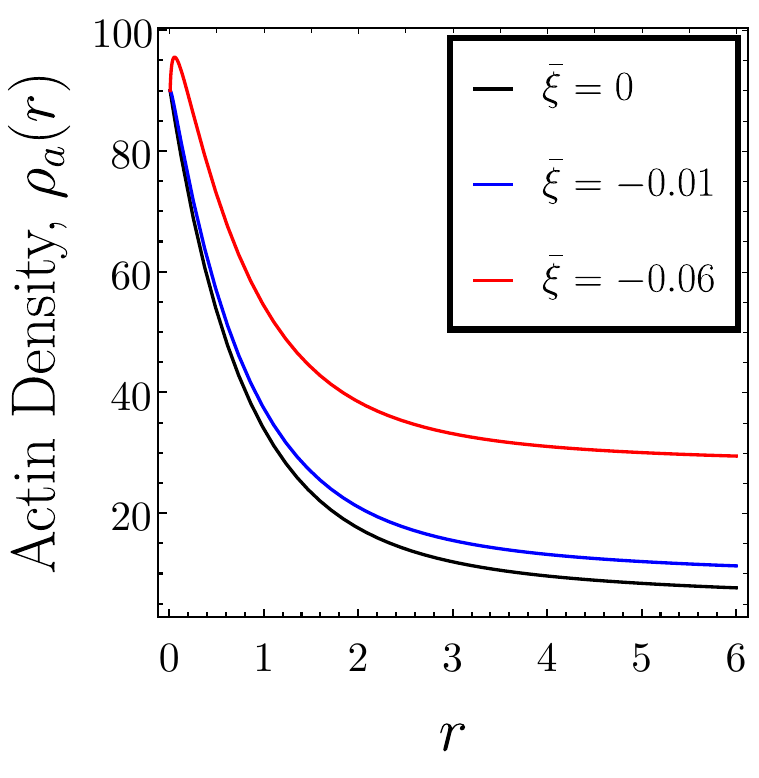}
\caption{\label{fig:neamticForces}Actin density in the case of the aster texture for different values of anisotropic active stress where $\bar{\zeta}=\frac{\bar{\xi}\rho_m}{k}$.}
\end{figure}

Here we will briefly justify our reasons for neglecting the anisotropic active part of the stress tensor in further analysis and in the main text. The anisotropic term is of the form
\begin{equation}
\sigma_{pp}= \bar\zeta \vec{P}\vec{P}
\end{equation}
depending on the sign of this term it can either stabilise or destabilise bend and splay respectively. This term would lead not only to an expulsion of actin from the centre of the aster (discussed in detail later), Fig.~\fig{fig:neamticForces}, but also to excessive shear banding not seen in experiments. For these reasons, we will choose to neglect it from our analysis, as it does not seem to play a large role in the system. Similar approaches in the literature also make this approximation; see, for example,  \cite{husain_emergent_2017}.

\subsection{Quasi-steady state equations for higher ATP concentrations: perturbative solution}
Here we outline the quasi-steady state multi-timescale expansion which we will use to find solutions for the spiral texture (we solve the full equations in the case of the aster, as they are genuine steady-state solutions at low ATP).

For low myosin density and high ATP concentration $\epsilon=\rho_m/k\rho_a$ is then a small parameter. Assuming the non-dimensionalised actin diffusion is of a similar order of magnitude to $\epsilon$ and adding the equations for myosin and actin force balance together gives
\begin{equation}
\vec{v}_a= \nabla\cdot\Sigma - \chi_a\nabla \rho_a \sim \mathcal{O}\left(\epsilon\right)\text{.}
\end{equation}

All terms in the stress tensor are of linear order in $\epsilon$ and $\kappa$ (as they all pre-multiply terms which go like the shear rate $u\sim \epsilon$ or $\vec{h}\sim\kappa$ and are thus quadratic in order).

The zeroth order equation for myosin force balance and first order for actin force balance are then given by
\begin{align}
&\vec{v}_m = -\vec{P} - \frac{\chi_m}{\rho_m} \nabla\rho_m + \mathcal{O}\left(\epsilon\right)\\
& \vec{v}_a = -\Bigg[\frac{1}{\rho_a}\nabla\cdot\Big\{\zeta(\rho_m,\rho_a)\mathbb{I}-\frac{1}{2}\left(\vec{P}\vec{h}-\vec{h}\vec{P}\right)\Big\}+\frac{\chi_a}{\rho_a}\nabla\rho_a\Bigg] \nonumber\\
&\qquad +\mathcal{O}\left(\epsilon^2\right)\text{.}
\end{align}

This system can then be solved hierarchically with solutions which go like
\begin{equation}
f(x)=\sum_n f^{(n)}(x)\text{ where }f^{(n)}\sim \mathcal{O}\left(\epsilon^r\kappa^s\right)\text{ and } r+s=n
\end{equation}
where we pick can pick a texture and initial actin density solve independently for the myosin velocity and density and use this to compute the actin velocity, and thus how the underlying texture remodels. Even at this level we can already see that this simple expansion gives an experimentally observable prediction in that the actin velocity at plentiful ATP will be much slower than the myosin velocity.

\subsection{Linear Stability Analysis}
Here we perform a linear stability analysis around an ordered uniform state to find a criterion where contractility sets in. We will assume that we are well into the nematic phase and that the actin density dependence of the myosin on rate has saturated.

We will assume an isotropic groundstate of the following form
\begin{align}
&\rho_m=\rho_m^{(0)} + \delta \rho_m=k^{-1} + \delta \rho_m\\
&\rho_a= \rho_a^{(0)} +\delta \rho_a\\
&P^i\vec{e}_i = \cos\phi \hat{e}_x +\sin\phi\hat{e}_y = \vec{e}_x + \delta\phi\hat{e}_y
\end{align}
where we have perturbed the $\phi=0$ orientation (i.e. all filaments pointing in the $x$ direction).

At lowest order the molecular field is given by
\begin{equation}
h^i=\Delta P^i\text{,}
\end{equation}
where $\Delta$ is the Laplacian. We write the contractility as
\begin{equation}
\zeta = \zeta^{(0)} +\zeta^{(1)}\delta\rho_m 
\end{equation}

The dynamical equations then read
\begin{widetext}
\begin{align}
&\delta \rho_{m,t}=\frac{\chi_{a}\rho_m^{(0)}}{\rho_a^{(0)}} \nabla^2\delta \rho_{a}+\left(\frac{\zeta^{(1)} \rho_m^{(0)}}{\rho_a^{(0)}} + \chi_m\right) \nabla^2\delta \rho_m - k\delta \rho_{m}  + \delta \rho_{m,x} + \rho_m^{(0)} \delta \phi_{,y}\\
&\delta \rho_{a,t}=\chi_a \nabla^2\delta \rho_{a}+\zeta^{(1)}\nabla^2\delta \rho_{m}\\
&\delta \phi_{,t}+\frac{\kappa}{4\rho_a^{(0)}}\Delta^2 \delta\phi = \kappa \Delta\delta\phi
\end{align}
\end{widetext}

Fourier transforming in space $\bar{f}(q) = \int\mathrm{d}^2x f(x)\exp(-i \vec{q}\cdot\vec{x})$ where $\vec{q}=(q_x,q_y)$ and $\vec{x}=(x,y)$, we find the following
\begin{equation}
\vec{X}_{,t} = A\cdot\vec{X}
\end{equation}
where $\vec{X}=(\delta\bar\rho_m,\delta\bar\rho_a,\delta\bar\phi)^\mathrm{T}$ and
\begin{widetext}
\begin{equation}
A= \left(
\begin{array}{ccc}
 -\frac{\zeta^{(1)} }{k\rho_a^{(0)}} |\vec{q}|^2-\chi_m |\vec{q}|^2+i q_x - k & -\frac{ \chi_a}{k\rho_a^{(0)}} |\vec{q}|^2 & i \frac{q_y}{k} \\
- \zeta^{(1)}|\vec{q}|^2 & -\chi_a|\vec{q}|^2 & 0 \\
 0 & 0 & -\frac{|\vec{q}|^2 \left(|\vec{q}|^2+4 \rho_{a}^{(0)}\right)}{4 \rho_{a}^{(0)}} \\
\end{array}
\right)
\end{equation}

We find the following eigenvalues for $A$
\begin{align}
&\lambda_1 =  -\frac{|\vec{q}|^2 \left(|\vec{q}|^2+4 \rho_{a}^{(0)}\right)}{4 \rho_{a}^{(0)}}\text{,}\\
&\lambda_2 = +\frac{\sqrt{\left(\left(\zeta^{(1)} \rho_m^{(0)} |\vec{q}|^2+\rho_a^{(0)} \left(|\vec{q}|^2 (\chi_a+\chi_m)-i q_x+k\right)\right)^2-4 (\rho_a^{(0)})^2 \chi_a |\vec{q}|^2 \left(\chi_{m} |\vec{q}|^2-i q_x+k\right)\right)}}{2 \rho_{a}^{(0)}}\nonumber\\
&-\frac{\zeta^{(1)} |\vec{q}|^2\rho_m^{(0)}}{2 \rho_{a}^{(0)}} + \left(|\vec{q}|^2 (\chi_a+\chi_m)-i q_x+k\right)\text{,}\\
&\lambda_3 = +\frac{\sqrt{\left(\left(\zeta^{(1)} \rho_m^{(0)} |\vec{q}|^2+\rho_a^{(0)} \left(|\vec{q}|^2 (\chi_a+\chi_m)-i q_x+k\right)\right)^2-4 (\rho_a^{(0)})^2 \chi_a |\vec{q}|^2 \left(\chi_{m} |\vec{q}|^2-i q_x+k\right)\right)}}{2 \rho_{a}^{(0)}}\nonumber\\
&-\frac{\zeta^{(1)} |\vec{q}|^2\rho_m^{(0)}}{2 \rho_{a}^{(0)}} - \left(|\vec{q}|^2 (\chi_a+\chi_m)-i q_x+k\right)\text{.}
\end{align}
In the low $q$ limit we find
\begin{align}
\{\lambda_1,\lambda_2,\lambda_3\}=\left\{-|\vec{q}|^2,-k+i q_x-\left(\frac{\zeta^{(1)}\rho_m^{(0)}}{\rho_a^{(0)}}+\chi_m\right)|\vec{q}|^2,-\chi_a |\vec{q}|^2\right\} + O\left(|q|^3\right)\text{,}
\end{align}
\end{widetext}
where writing $|q|=Q$, $\rho_m^{(0)}=k^{-1}$ and $\zeta^{(1)}=\frac{\xi}{k}$ allows us to find the following instability condition for $\lambda_2$
\begin{equation}
    0\leq -k-\left(\frac{\xi}{k^2\rho_a^{(0)}}+\chi_m\right)Q^2\text{.}
\end{equation}

\begin{figure}
    \centering
    \includegraphics[width=0.4\textwidth]{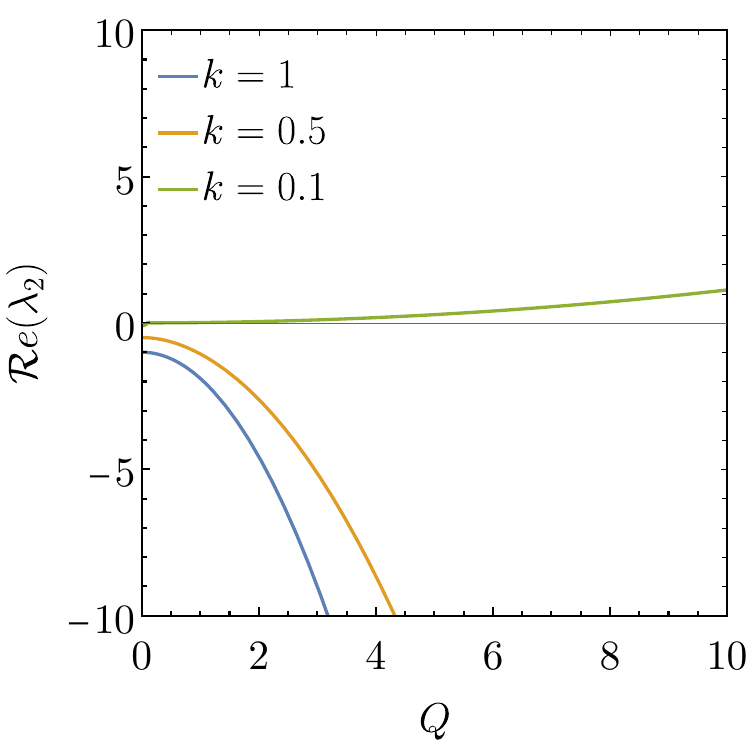}
    \caption{Real part of the second eigenvalue for $\xi=-1$, $\rho_a^{(0)}=10$ and $\chi_m=1$ for a selection of $k$ values.}
    \label{fig:siStabilityPlots}
\end{figure}

We plot the full growth rate of $\lambda_2$ in Fig.~\fig{fig:siStabilityPlots} along with a phase diagram compared to the case where $\zeta^{(1)}=\text{Const}$ (where the constant is chosen to match the phase diagram in the high $Q$ regime), Fig.~\ref{fig:numericsFig}a.

\subsection{Full Numerical Solutions}\label{sec:fullNumerics}
To solve the full hydrodynamic equations numerically we use a $100\times 100$ periodic grid with $\Delta x=0.1$ and use a central difference stencil in space. To integrate in time, we use a $4^{\text{th}}$ order Runge-Kutta method with $\Delta t=0.005$ such that the numerical integration is stable (assuming a standard von Neumann numerical stability condition for reaction diffusion equations). We use bespoke code in C++ to solve this numerically and use PyPlot to create the figures and videos using a script based on openCV functions. We choose a uniform density initial condition on actin $\rho_a(0)=10$, myosin $\rho_m(0)=1/k$ and a uniform random orientation on the initial polarisation $\vec{P} = \cos(\theta)\vec{e}_x+\sin(\theta)\vec{e}_y$ where $\theta\sim \mathcal{U}[0,2\pi)$.

\subsection{Numerical solution to the steady state equations}
Here we discuss the numerical solutions in the case of the ``aster'' and ``spiral'' textures.

\subsubsection{Aster/contractile foci texture}
For the regime with mostly depleted ATP we consider an aster texture as the possible steady state. An axisymmetric steady state texture of the following form is chosen
\begin{equation}
\vec{P}=\vec{e}_r;\quad \psi=0\text{.}
\end{equation}
This give $\vec{h}=0$ as the texture has no bend and actin velocities are also zero by symmetry arguments, conservation of actin mass and a little algebra.\footnote{We can argue that on physical grounds the aster is stable as the anisotropic active forcing or elastic splay forcing needed to destabilise them are not present in our model. The fact that we see such stable structures in an experiment is a further justification for neglecting these terms}.
 
The force balance equations for myosin (in the absence of actin flows, which are zero for the aster texture) now reduce to the following equations for the myosin velocity
\begin{equation}
v_m^r = -\frac{\chi_m \partial_r\rho_m}{\rho_m}-1;\quad v_m^\theta = 0\text{.}
\end{equation}
This implies that $\vec{v}_m$ is a function of $\rho_m$, such that the accumulation of myosin-II minifilaments due to their procession towards texture's centre is balanced by the density dependent dissociation rate.

This can be shown more directly by substituting into the continuity equation for myosin to give
\begin{equation}
\frac{\chi_m r \partial_r^2\rho_m+(\chi_m+r) \partial_r\rho_m + (1-k r) \rho_m+r}{r}=0
\end{equation}
which is a boundary value problem for $\rho_m(r)$ with boundary conditions of no flux at the origin and no gradients at infinity
\begin{equation}
\rho_m(0)=-\chi_m\partial_r\rho_m|_{r=0};\quad \partial_r\rho_m|_{r\to \infty}=0\text{.}
\end{equation}

Neither of these boundary conditions can be implemented exactly in a numerical scheme, so we choose to apply the first boundary condition at a small but finite radius, choosing a value of $\rho_m(0)$ which we then use as a shooting parameter to find a solution which is converging to flat in the far field.

With this solution, we can now proceed to find the actin density by substituting into the remaining force balance condition for actin, which equates contractile forces with those resisting the compression of actin.  This gives the following first order equation for $\rho_a$ that is readily solvable numerically
\begin{equation}
\frac{\chi_a k r \partial_r\rho_a + r (\bar\xi+\xi) \partial_r\rho_m + \bar\xi \rho_m}{k r}=0\text{.}
\end{equation}
Here we keep the most general form, including anisotropic terms. In the main paper, we set these to zero.
We solve this for a boundary condition at $r=0.01$ that is approximately in agreement with the actin and myosin density ratios seen in the experiment ($\sim 10:1$).

We give details of all parameters used for this case in Appendix D9, where we also detail the fitting to data for the two clearest structures in our data-set.

\subsubsection{Spiral/vortex texture}
We now turn our attention the the quasi-steady state solutions that have a spiral/vortex-like character. In order to explain the long-lived vortex / spiral motifs of our experiments, we must take a perturbative approach: expanding in small $\epsilon$ (high ATP) and solving hierarchically. At zeroth order, $\vec{v}^{(0)}_a = 0$, implying freedom to impose $\vec{P}^{(0)}$ and $\rho_a^{(0)}$. We pick the following texture for the spiral in order to mimic the actin seen in TIRF and the flow dynamics of the myosin seen in iSCAT
\begin{equation}
\psi^{(0)} = \frac{\pi}{2}\frac{r^2}{b+r^2}\text{.}
\end{equation}
This texture is plotted as an inset in the main text, Main Text Fig.~3f. We also choose the actin density profile $\rho_a^{(0)} = c r^2 e^{-r^2/R^2} +l$ with parameters to mimic a peak in actin density in the ring (as seen in TIRF data).

Making use of the multi-timescale expansion which is valid at higher ATP concentrations where these structures are seen simplifies the equations considerably and means that we can solve the myosin flows on a fixed actin texture before computing the first-order corrections to the actin flow later. The myosin mass conservation equation is given by
\begin{equation}\label{eq:myoContVortex}
k-1+ \rho_m \partial_rv_m^r + v_m^r \left(\partial_r\rho_m + \frac{\rho_m}{r}\right) = 0
\end{equation}
and the myosin force balance by
\begin{align}
\vec{e}_r:\quad  \rho_\text{m} \left(\sin \left(\frac{\pi  b}{2 \left(b+r^2\right)}\right)+v_\text{m}^r\right) &\nonumber\\
\qquad +\chi_\text{m} \partial_r\rho_\text{m} &=0\\
\vec{e}_\theta:\quad  \rho_\text{m} \left(\cos \left(\frac{\pi  b}{2 \left(b+r^2\right)}\right)+v_\text{m}^\theta\right) &= 0\text{.}
\end{align}
In this case, the minifilament velocity profile, $\vec{v}_m$ is that which balances propulsive forces with density gradients.

Rearranging and substituting these expressions for the myosin velocity in terms of the myosin density into the continuity equations (Eq.~\eq{eq:myoContVortex}) this can be solved using the same shooting method as in the aster case, as to lowest order in $\epsilon$ these equations are completely decoupled from the actin dynamics. The boundary conditions are identical to the aster case and we can solve again numerically with a shooting method.

\subsubsection*{First order corrections to the actin velocity}
As might be expected, this state is unstable at first order in $\epsilon$.  We can, nevertheless, calculate first order corrections to the actin velocity profile, $\vec{v}_a^{(1)}$. As stated earlier we choose an initial actin density peaked about the ring (in imitation of the densities seen in TIRF experiments), hence we choose a ground state density of the the following form (for the parameters used to numerically solve the myosin dynamics previously)
\begin{equation}
\rho^{(0)}_a(r) = c r^2 e^{-r^2/R^2} + l\text{.}
\end{equation}

The actin force balance equation then reads
\begin{align}
\vec{e}_r:\quad &\frac{-\chi_\text{a} k \partial_r\rho_\text{a}- \xi  \partial_r\rho_\text{m}}{k \rho_\text{a}}=v^{r}_\text{a}\\
\vec{e}_\theta:\quad & -\frac{4 \pi  b \kappa  r \left(2 b-r^2\right)}{\left(b+r^2\right)^4 \rho_\text{a}}=v^{\theta}_a
\end{align}
which gives the actin velocity $\vec{v}_a^{(1)}$.

\subsection{Comparison to data}
\subsubsection{Axisymmetric averaging of the iSCAT data}
Plots of the axisymmetric density of myosin in actomyosin vortices, spirals and asters were obtained from maximum projections of 300 frames (time series of 1 min) using Fiji and the Radial Profile Angle plug in (\href{https://imagej.net/ij/plugins/radial-profile-ext.html}{https://imagej.net/ij/plugins/radial-profile-ext.html}). The maximum radius was chosen to cover the size of the feature (\(3.5 - 5.5 \mu m \) and the normalised integrated intensity was calculated for the full circumference of each increment in radius.

\subsubsection{Calculation of radial and tangential velocities from PIV lab data}
Using PIV lab analysis, we obtained velocity vector maps of vortices, spirals, and asters that are represented as a grid in Cartesian coordinates with velocity vectors ($v_x, v_y$). The central position of the actomyosin features, ($x_c, y_c$), was manually determined. For each point on the grid, the relative position vector from the centre to that point ($x_{rel}, y_{rel}$) was calculated. The radial and tangential components of the velocity ($v_r$ and $v_t$ respectively) was calculated using the following formula: 
\begin{align*}
&v_r = \frac{ x_{\text{rel}} \cdot v_x + y_{\text{rel}} \cdot v_y}{\sqrt{ x_{\text{rel}}^2 + y_{\text{rel}}^2}} \\
&v_t = \frac{y_{\text{rel}} \cdot v_x -  x_{\text{rel}} \cdot v_y}{\sqrt{ x_{\text{rel}}^2 + y_{\text{rel}}^2}}
\end{align*}
To average the velocities at a certain radius from the center, reference radii were defined as $r = [0, 0.5, 1.0, 1.5, \ldots, 6.0] \, \mu m$, resulting in $N = 12$ circles around the center. For each radius, we defined a band of thickness $0.2 \, \mu m$ in which the velocities were averaged, i.e. we calculated the mean velocities for the regions $r_i$ to $r_i + 0.2 \, \mu m$, where $i$ is an integer ranging from 1 to 13.

\begin{figure}[t]
    \centering
    \includegraphics[width=\linewidth]{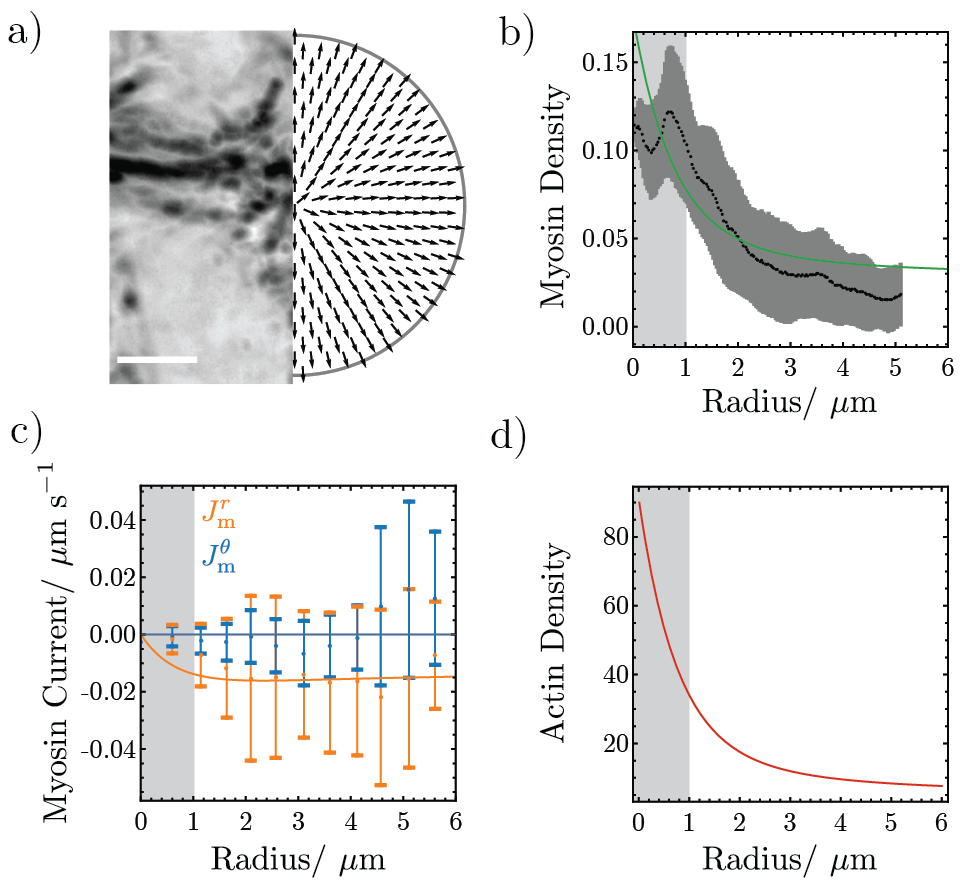}
    \caption{\textbf{a)}Left: Static iSCAT microscopy image of an aster formed after 15 minutes in the intermediate ATP regime.
  \textbf{a)}Right: Underlying asterlike actin texture for the steady-state solution.
  \textbf{ b)} The radial myosin density profile compares well with the angular average of the interferometric scatting intensity.
\textbf{c)} The components of the myosin current broadly agree with density weighted velocities from a PIV analysis.
\textbf{d)} The radial actin density profile (not captured by iSCAT) mirrors the functional form of the myosin density profile.}
    \label{fig:asterHighATP}
\end{figure}

\subsubsection{Fitting procedure}
We select the two clearest textures in the iSCAT data at high and low ATP concentrations to compare with the axisymmetric theory. These are the aster, SM Fig. 2, and vortex ring, SM Fig. 3 (far right) in the supplementary material \cite{supp_al-izzi}. In both we rescale the velocity to the maximum myosin velocity in the averaged data. 

For the aster this directly gives us the results in Fig.~\ref{fig:hydrodynamicsFig}(a-d) with parameters $\xi=-1$, $\chi_m=1$, $\chi_a=0.1$ and $k=0.8$. We rescale the density to the iSCAT intensity by minimising the $L_2$ norm between the the density and the iSCAT intensity data for $r\geq 1\mu\text{m}$.

For the vortex we fit the parameters $c$, $l$ and $b$ to the velocity data by minimising the $L_2$ norm. We pick values for $R=3.4\mu\text{m}$ and $\rho_a^\text{h}=5$ to match the the iSCAT secondary peak at $r=3.4\mu\text{m}$ (Fig.\ref{fig:hydrodynamicsFig}f) and to restrict the number of parameters searched for (the resulting values of $c$, $l$ and $b$ were insensitive to the value of $\rho_a^h\sim 1$). We find values of $c=0.5$, $l=0.1$ and $b=0.05$ respectively with a risidual of $\text{error}^2=0.0746603$. We then rescale the density to fit the iSCAT intensity data as in the aster case.

\subsubsection{Additional fit to aster texture in High ATP regime}
Here we show an equivalent plot for an aster texture but observed at the same levels of ATP as the vortex structure in Fig.~\ref{fig:hydrodynamicsFig}(e-h). This lends further support to asters being the true steady state of the system and being possible at both high and low ATP.

\bibliography{bibliography}

\begin{thebibliography}{70}
\bibitem [{sup()}]{supp_al-izzi}%
  \BibitemOpen
  \href@noop {} {\enquote {\bibinfo {title} {Electronic supplamentary material
  can be found at: ...... detailing additional figures, experimental details \&
  calculations.}}\ }\BibitemShut {NoStop}%
\end{thebibliography}

\end{document}